# Sustainability-Aware Discovery of Dual-Atom Hydrogen Evolution Catalysts via a Unified Machine Learning Framework for Activity and Carbon Cost

*Yanmei Zang[1†*], Hyun Gyu Park[1†], Gi Beom Sim[1], Tae Hyeon Park[1], Ho Jin Lee[1], Xiaorong Zou[2], D. ChangMo Yang[1], Soohaeng Yoo Willow[1], Hye Jung Kim[2*], Sunbin Yoo[1,3,4]* Chang Woo Myung[1,2,3,4*]*

[1]Department of Energy Science, Sungkyunkwan University, 2066 Seobu-ro, Suwon 16419, Republic of Korea

[2]Center for 2D Quantum Heterostructures (2DQH), Institute for Basic Science (IBS), Suwon 16419, Republic of Korea

[3]Department of Energy, Sungkyunkwan University, Seobu-ro 2066, Suwon 16419, Republic of Korea

[4]SKKU National Lab for Intelligent Energy Solution Technology (SIEST), Sungkyunkwan University, Seobu-ro 2066, Suwon 16419, Republic of Korea

*E-mail: yanmei@skku.edu (Y.Z.); hjkim75@skku.edu (H. J. K.); sunbinyoo@skku.edu (S. Y.); cwmyung@skku.edu (C. W. M.)

†Contributed equally to this work

## Abstract

Catalyst discovery for green hydrogen has optimized activity alone, deferring economic and environmental costs to stages where they disqualify otherwise promising candidates. Here we introduce a unified machine learning framework that screens catalytic activity and monetized carbon cost in a single pipeline. A machine learning potential predicts hydrogen evolution activity and stability at density-functional-theory accuracy. A second model, trained on published estimates of the social cost of carbon, converts each candidate's embodied $CO_2$ into a monetary burden. Screening 460 graphene-supported dual-atom catalysts across 23 transition metals and 20 nitrogen-coordination motifs yields 26 active, synthesizable candidates. The two axes then conflict. The most active candidates span Mn-, Fe-, Zr-, Cr-, Cu-, Ni-, Ti-, Nb-, V-, Pd-, Au-, and Rh-based catalysts, indistinguishable in adsorption energy yet spanning a 58,000-fold range in carbon cost. Joint optimization selects $Mn_2@2N_f$ and $Zr_2@2N_c$, matching the best precious-metal activity at tens-of-thousands-fold lower carbon cost, independent of catalyst loading.

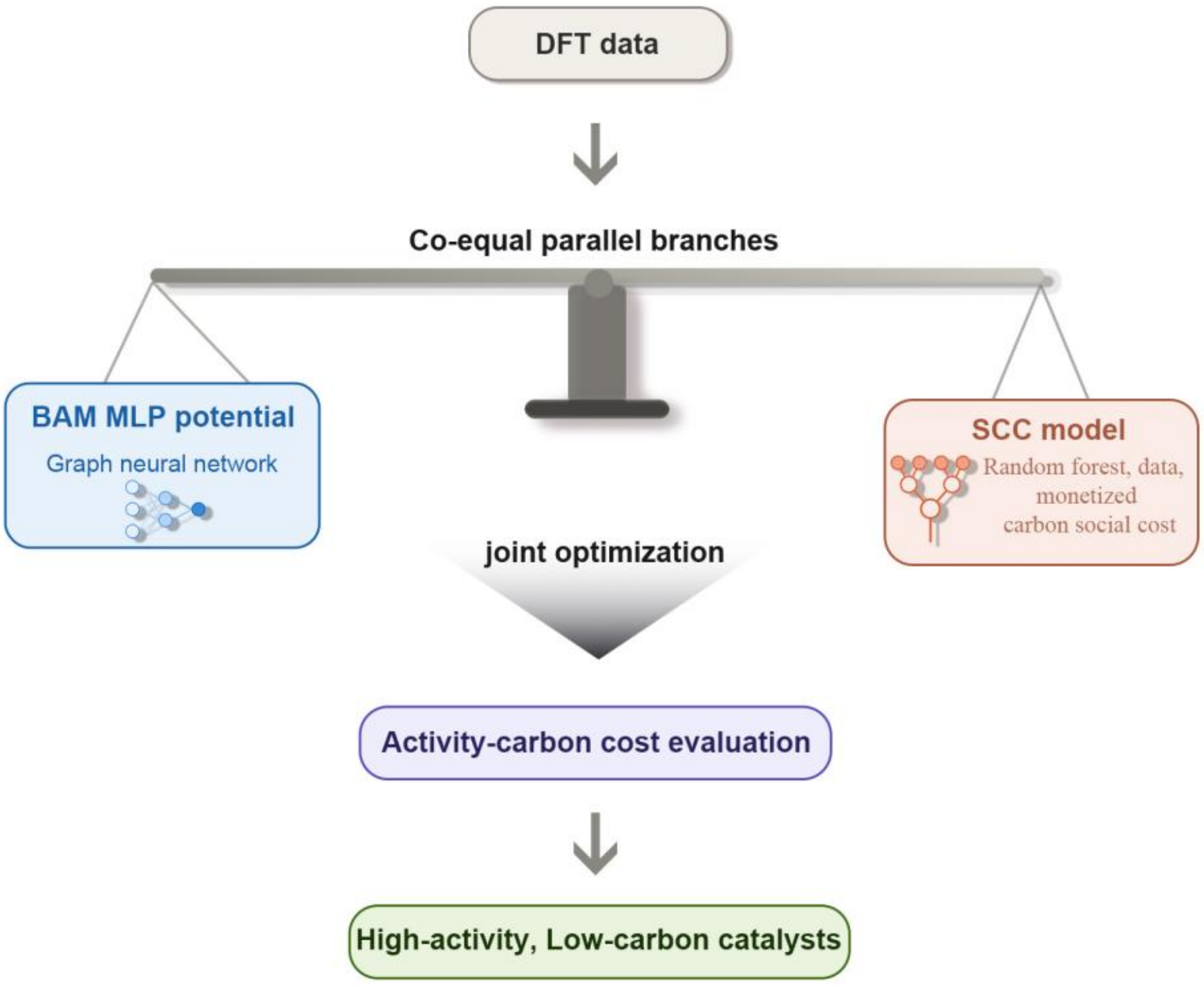

This work establishes a unified framework that couples machine learning potential-based catalyst screening with carbon social cost assessment, enabling catalytic activity and climate-related environmental cost to be optimized simultaneously.

## 1. Introduction

The rational design of electrocatalysts is central to advancing sustainable energy technologies, as the atomic-scale structural and electronic properties of the active sites intimately govern catalytic performance.[1-3] For the hydrogen evolution reaction (HER), decades of experimental and theoretical research have established that small variations in local coordination environment, metal identity, and support interaction can lead to dramatic differences in catalytic activity, selectivity, and stability.[4-10] Among the many candidate platforms, graphene-supported dual-atom catalysts (DACs) have attracted growing attention because they maximize metal utilization, offer tunable coordination environments, and enable strong metal–support interactions that collectively allow precise modulation of hydrogen adsorption energetics.[11-17] However, the enormous combinatorial space defined by transition-metal identity, metal pairing, and nitrogen-coordination motif renders exhaustive experimental exploration impractical, motivating the development of efficient computational screening strategies.

Density functional theory (DFT) has long been the cornerstone of computational catalyst design, providing reliable predictions of adsorption energetics, electronic structure, and reaction pathways at the atomic scale.[18-21] High-throughput DFT screening has enabled the identification of promising HER catalyst candidates, with hydrogen adsorption free energy ($\Delta G_{H*}$) established as the primary activity descriptor.[22,23] However, DFT-based screening faces inherent limitations: its computational cost scales poorly with system size and configurational complexity, making exhaustive exploration of large DAC libraries computationally prohibitive. As a result, most DFT studies remain limited to a narrow subset of metal combinations and coordination motifs, leaving broad regions of the catalyst design space underexplored. More importantly, even when DFT screening successfully identifies promising candidates, it optimizes exclusively for activity descriptors, systematically neglecting the environmental costs intrinsic to catalyst production.[24-26] And this limitation grows increasingly untenable as sustainability considerations move to the forefront of materials discovery.

Machine learning (ML) has emerged as a transformative approach to bridge this gap, enabling rapid and accurate prediction of catalytic properties across vast chemical spaces at a fraction of the DFT cost.[27-29] ML potentials and surrogate models trained on DFT datasets have demonstrated the ability to predict adsorption energetics, Gibbs free energy, and stability with near-DFT accuracy, opening the door to truly high-throughput catalyst screening.[30-36] Nevertheless, a critical and previously unaddressed challenge remains: computational catalyst screens have consistently prioritized catalytic activity alone, while the upstream carbon burden associated with metal production has never been incorporated as a co-equal, quantitative criterion within the same screening pipeline.[25,26] This omission is consequential that catalysts based on precious or carbon-intensive metals may appear highly promising when solely by activity descriptors, yet their synthesis can generate orders of magnitude more $CO_2$ than earth-abundant alternatives, fundamentally undermining their sustainability credentials.

Whether a catalyst is ultimately deployed depends as much on economics as on activity. Techno-economic assessment (TEA) and life-cycle assessment (LCA) have therefore become instruments for evaluating hydrogen production routes, translating catalyst loading, precious-metal content, durability, and electricity demand into a levelized cost of hydrogen and a cradle-to-gate carbon footprint.[37-41] Applied to water electrolysis, these assessments identify element scarcity and the embodied emissions of metal production, rather than intrinsic activity, as the binding constraints on terawatt-scale deployment.[26,42-44] Carbon cost adds a dimension that neither capital cost nor tonnes of $CO_2$ captures on its own. The social cost of carbon (SCC) monetizes the long-run climate damage caused by one additional tonne of $CO_2$, so that an environmental burden can be expressed in the same unit as material and electricity costs, compared across technologies, and entered directly into cost-benefit analysis, carbon pricing, and investment appraisal.[45-48] In materials discovery, however, such evaluations are almost always performed ex post, for one or a few candidates after synthesis, when they can no longer alter which candidates were screened in the first place.[49-51] Machine learning lowers the cost of moving these evaluations upstream. Surrogate models trained on life-cycle inventory databases now predict cradle-to-gate impacts of chemicals and materials directly from compositional descriptors,[52,53] and, in climate economics, ML meta-analysis synthesizes thousands of heterogeneous SCC estimates while isolating the structural choices, chiefly the discount rate and the persistence of damages, that drive them.[48,54] An economic or carbon criterion can therefore be evaluated at the same throughput, and from the same inputs, as an activity descriptor, which is what allows the two to be optimized jointly rather than sequentially.

To address this critical gap, we introduce a unified machine learning framework that integrates the social cost of carbon (SCC) as a quantitative, co-equal optimization criterion alongside catalytic activity in a single HER catalyst discovery pipeline. The social cost of carbon, defined as a monetized estimate of the long-term economic damage incurred by $CO_2$ emissions, provides a principled and quantitative framework for embedding environmental cost as a first-class criterion in materials screening workflows. Yet, efforts to jointly consider catalytic activity and environmental cost within a single computational catalyst design workflow have remained relatively scarce.[55-57] In our framework, a BAM machine learning potential (MLP)[58] trained on DFT reference data predicts HER-relevant physical descriptors, including binding energy and $\Delta G_{H*}$ across a library of 460 graphene-supported $TM_2@N_x$–Gr DACs spanning 23 transition metals and 20 nitrogen-coordination environments. A linked carbon-cost module then converts each candidate's composition into a monetized social cost of carbon using an ML-derived SCC model, enabling the joint ranking of catalysts by both activity and embodied carbon burden within a single, seamless pipeline. Our results reveal a pronounced disconnect between the two objectives: candidates that are effectively indistinguishable in catalytic performance nonetheless span a 58,000-fold range in associated carbon cost. Joint optimization identifies Mn- and Zr-based DACs that match the catalytic activity of the best precious-metal catalysts at tens-of-thousands-fold lower social cost of carbon demonstrating that sustainability-aware ML screening can effectively redirect catalyst discovery toward earth-abundant, environmentally responsible materials without any compromise in performance.

## 2. Results and discussion

### 2.1 Joint Machine Learning and Carbon-Cost Screening Framework and $TM_2@N_x$–Gr Catalyst Library

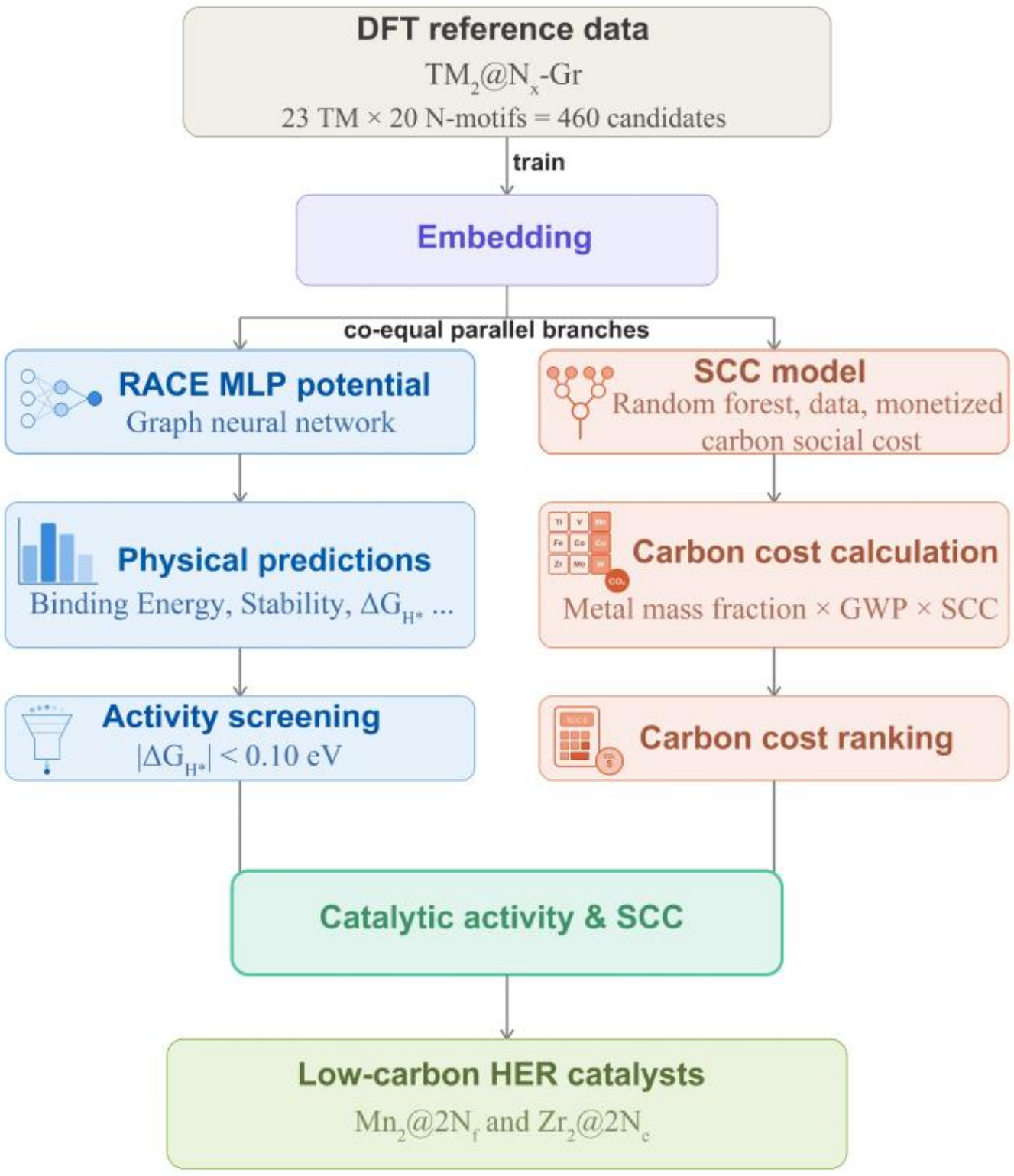

**Figure 1.** Schematic overview of the joint machine learning and carbon-cost screening framework for HER catalyst discovery.

**Figure 1** illustrates the unified machine learning and carbon-cost screening framework developed in this work. The workflow begins with a DFT reference dataset of 460 $TM_2@N_x$–Gr candidates spanning 23 transition metals and 20 nitrogen-coordination motifs, from which a shared embedding layer encodes the structural and chemical information of the full catalyst library. Details of the nitrogen coordination motifs and transition-metal set are provided in **Figure S1**. From this common embedding, the framework branches into two parallel pipelines. The left branch deploys a BAM machine learning potential based on a graph neural network to predict HER-relevant physical descriptors (including binding energy, stability, and $\Delta G_{H^*}$) and applies an activity threshold of $|\Delta G_{H^*}| < 0.10$ eV to identify catalytically promising candidates. The right branch employs a random forest model trained on published SCC estimates to quantify the embodied carbon burden of each candidate, computing carbon social cost as the product of metal mass fraction, metal-specific global warming potential (GWP), and the monetized SCC. Although the two branches are applied in sequence, they carry equal weight in the final candidate ranking: carbon social cost is not a secondary filter but a co-equal optimization criterion that reshapes catalyst prioritization as fundamentally as catalytic activity itself. The outputs of both branches are merged into a joint activity-sustainability ranking, from which the final low-carbon HER catalysts ($Mn_2@2N_f$ and $Zr_2@2N_c$) are identified as candidates that simultaneously achieve near-optimal activity and minimal embodied carbon cost.

## 2.2 Thermodynamic Stability Screening across Coordination Environments

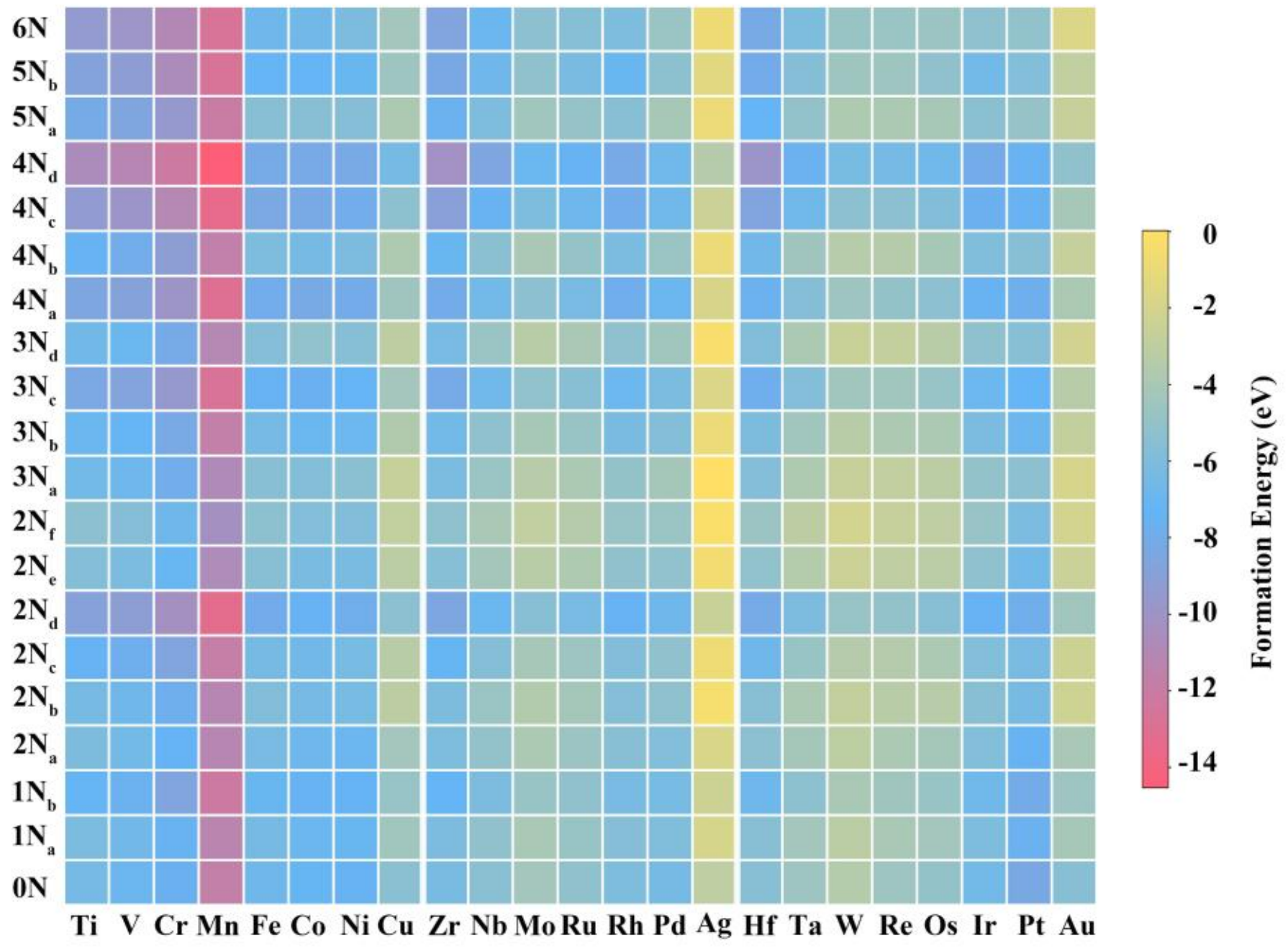

**Figure 2.** Formation energy map of $TM_2@N_x$–Gr catalysts across the screened design space. The horizontal axis lists the 23 transition metals spanning the 3d (Ti–Cu), 4d (Zr–Ag), and 5d (Hf–Au) series, and the vertical axis lists the 20 nitrogen coordination environments from pristine double vacancy (0N) through 1N, 2N, 3N, 4N, and 5N motifs (with structural isomers denoted by subscripts a–f) to fully coordinated 6N. The color scale encodes the formation

energy $E_f$ (in eV), defined in Eq. (3), ranging from approximately −14 eV (most stable, red) to 0 eV (least stable, yellow). Numerical values for all 460 configurations are provided in **Table S1.**

Having established the joint screening framework, we first evaluate the thermodynamic stability of the full $TM_2@N_x$–Gr library, as stability is a prerequisite for any candidate to advance through both the activity and carbon-cost branches of the workflow. The formation energies of all 460 coordination configurations were systematically analyzed, as summarized in **Figure 2** and **Table S1**. Overall, the majority of $TM_2@N_x$–Gr systems exhibit negative formation energies, indicating that the incorporation of diatomic metal species into N-doped graphene is thermodynamically favorable. Such stabilization is particularly important for suppressing metal aggregation and improving the structural durability of dual-atom catalysts under electrochemical conditions. As shown in **Figure 2**, the formation energies depend on both the nitrogen coordination number and the local geometric configuration around the metal dimers. In general, increasing nitrogen coordination significantly enhances the thermodynamic stability of the $TM_2$ centers, and intermediate coordination environments (2N–4N) generally exhibit the lowest formation energies, suggesting that moderate nitrogen coordination provides an optimal balance between structural stabilization and local electronic flexibility. Formation energies also vary systematically across different transition metals: early transition metals generally exhibit more negative values than late transition metals, reflecting stronger metal–nitrogen bonding driven by differences in d-electron occupation. Notably, Mn-containing systems exhibit particularly low formation energies across nearly all coordination environments, implying exceptionally strong stabilization of Mn dimers within N-doped graphene frameworks. This trend is further corroborated by the subsequent screening results, which identify Mn-based candidates as promising HER catalysts with low carbon costs. In contrast, noble-metal systems such as Pt and Au tend to show comparatively less negative formation energies, reflecting weaker metal–support interactions. Furthermore, even under the same nitrogen coordination number, different local coordination geometries lead to noticeable variations in formation energy. For $Ti_2@N_a$, $Ti_2@N_b$, $Ti_2@N_c$, $Ti_2@N_d$, and $Ti_2@N_e$, the formation energies are −6.03, −6.38, −7.39, −8.85, and −5.74 eV, respectively. This geometric sensitivity originates from differences in local symmetry, metal–metal distance, and orbital hybridization between the transition-metal dimers and neighboring nitrogen atoms, and underscores that coordination number alone is an insufficient stability descriptor. Overall, these results show that nitrogen coordination engineering can effectively tune the thermodynamic stability of $TM_2@N_x$–Gr

catalysts, and formation energy is carried forward as the stability screening criterion alongside the HER activity and electronic-structure descriptors evaluated in the following sections.

### 2.3 Validation of the BAM MLP for Physical Descriptor Prediction

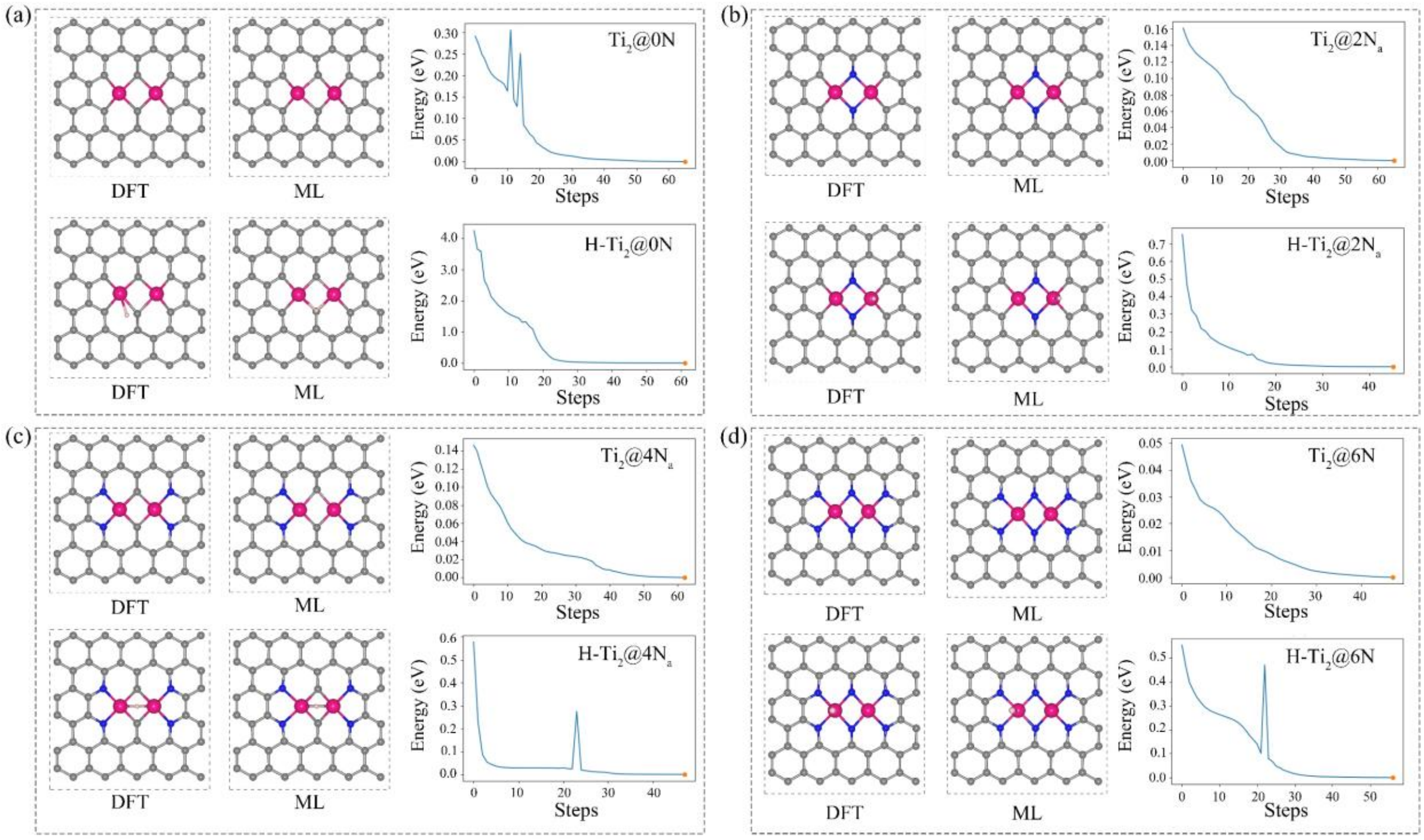

**Figure 3.** The optimized structures obtained by the MLP are compared with the DFT-relaxed geometries, while the corresponding energy evolution profiles are plotted as a function of geometry optimization steps for representative $TM_2@N_x$-Gr and H–$TM_2@N_x$-Gr systems: (a) $Ti_2@0N$, (b) $Ti_2@2N_a$, (c) $Ti_2@4N_a$, and (d) $Ti_2@6N$.

Having established the thermodynamic stability landscape of the $TM_2@N_x$–Gr library, we next validate the BAM MLP that drives the physical branch of the joint screening framework. This validation step is critical not only for confirming the accuracy of activity predictions, but also for the integrity of the overall activity–carbon-cost ranking: a joint optimization framework is only as meaningful as the reliability of each of its constituent evaluation branches. As shown in **Figure 3**, four representative configurations – $Ti_2@0N$, $Ti_2@2N_a$, $Ti_2@4N_a$, and $Ti_2@6N$ – were selected to assess the structural optimization performance of the trained model for both $TM_2@N_x$–Gr and H–$TM_2@N_x$–Gr systems. The optimized structures obtained from the BAM MLP exhibit excellent agreement with the corresponding DFT-relaxed geometries, demonstrating that the trained model accurately captures the local atomic configurations of both dual-atom active centers and hydrogen-adsorbed intermediates across different nitrogen coordination environments. The corresponding energy evolution profiles further confirm the stability of BAM MLP geometry optimization: in all representative cases, the total energy decreases rapidly in the initial steps and converges smoothly toward a minimum, with no signs of divergence or unphysical energy fluctuations. This well-behaved convergence is essential for reliable $\Delta G_{H^*}$ prediction across the 460-candidate library, as erratic optimization trajectories would introduce systematic errors into the activity descriptors that feed the downstream carbon-cost ranking. These validation results confirm that the BAM MLP delivers sufficient accuracy for high-throughput physical screening, providing the reliable activity predictions that meaningful integration with carbon social cost evaluation demands.

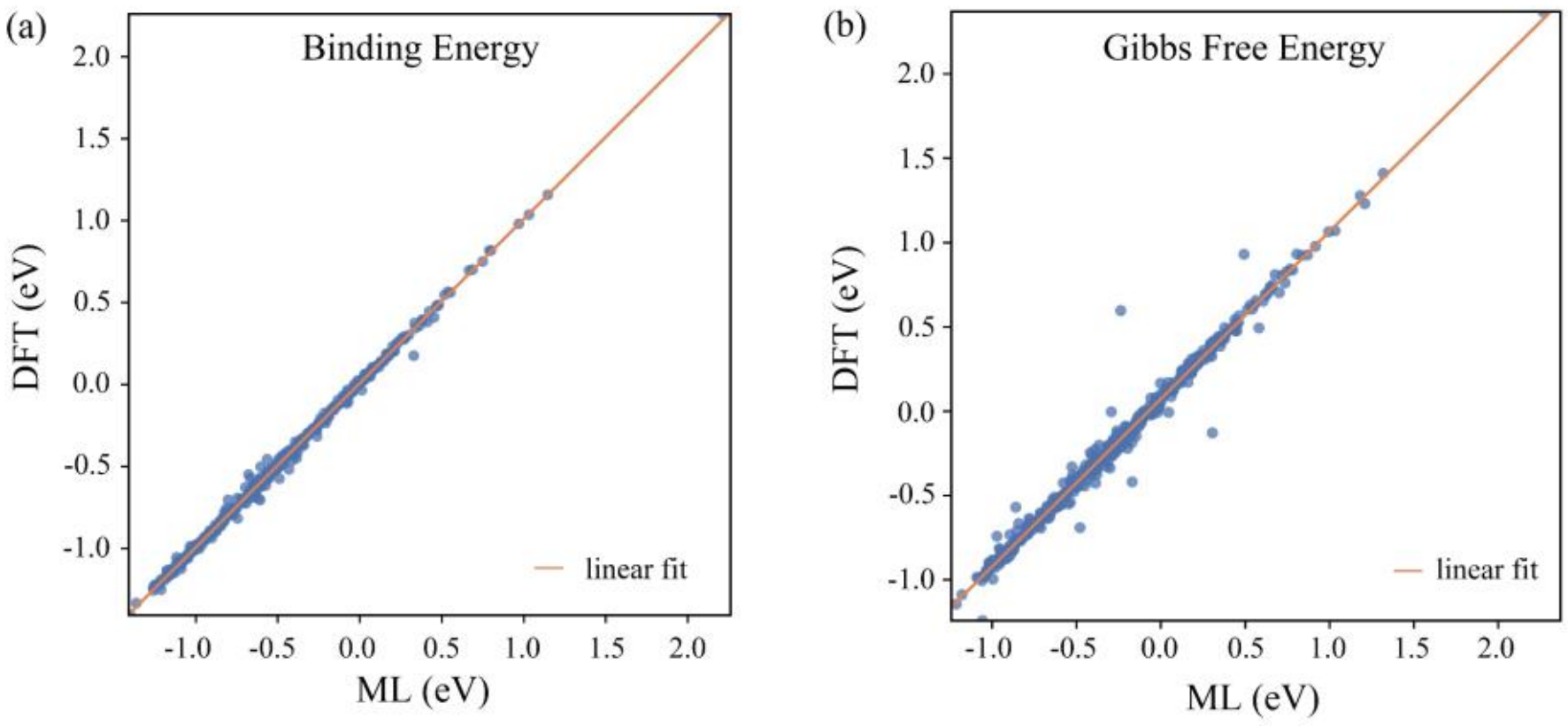

**Figure 4.** Correlation between ML predictions and DFT calculations for (a) binding energy and (b) Gibbs free energy. Each point represents an individual $TM_2@N_x$–Gr catalyst. The solid orange line denotes the linear fit to the data.

To further evaluate the reliability of the BAM MLP, we systematically compare its predictions with DFT calculations for both binding energy and Gibbs free energy across the full $TM_2@N_x$–Gr library (**Figure 4**). As shown in **Figure 4(a)**, the predicted binding energies exhibit a strong linear correlation with DFT results, with a high $R^2$ value of 0.997 and very low errors (mean absolute error (MAE) = 17 meV and root mean square error (RMSE) = 25 meV), confirming that the trained model accurately capture the adsorption energetics across different catalytic systems.

The Gibbs free energy predictions shown in **Figure 4(b)** also correlate strongly with DFT results, yielding an $R^2$ of 0.961, MAE of 80 meV, and RMSE of 97 meV. The slightly larger deviations compared with binding energy predictions can be attributed to the additional thermodynamic corrections, including zero-point energy and entropy contributions.

Nevertheless, the data points remain closely distributed around the ideal linear fit, further confirming the robustness of the model across the full chemical space screened.

Critically, both error metrics fall well within the accepted threshold for screening-level applications: prediction errors below ~100 meV are widely considered sufficient for identifying promising catalyst candidates and evaluating activity trends, where relative energetic ordering matters more than absolute numerical precision. This level of accuracy is particularly consequential in the present framework, where $\Delta G_{H*}$ predictions from the BAM MLP directly determine which candidates advance to carbon-cost evaluation — an erroneous activity ranking would propagate into the joint activity–sustainability ranking and ultimately misidentify the low-carbon HER catalysts. The validated BAM MLP thus provides the accurate and efficient physical screening foundation that rigorous integration with the carbon social cost evaluation requires.

## 2.4 HER Activity Screening and Candidate Selection for Carbon-Cost Evaluation

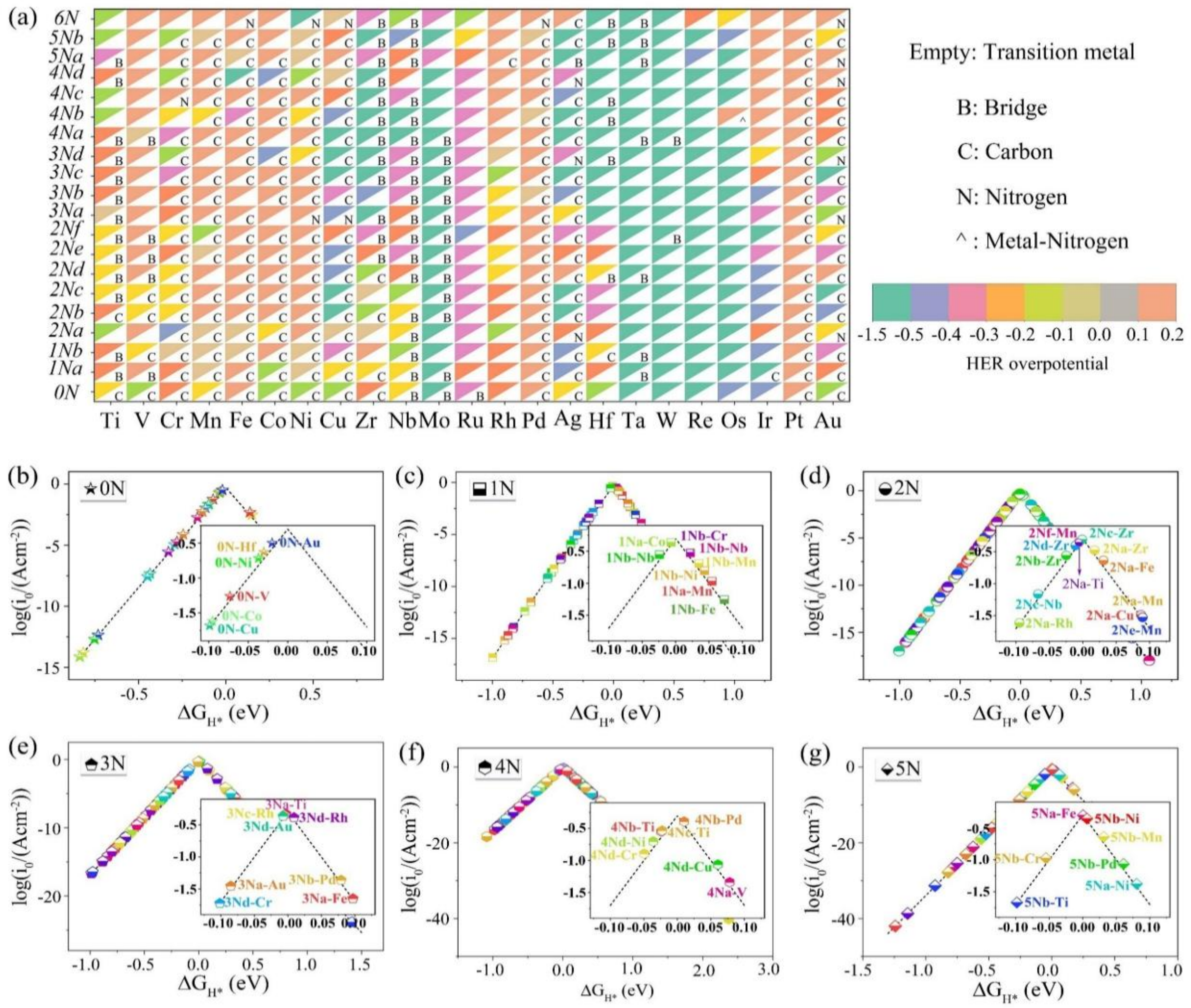

**Figure 5.** (a) Triangular heat map summarizing the HER overpotential ($\eta_{HER}$, upper triangle) and the corresponding most stable hydrogen adsorption configurations (H*, lower triangle). The adsorption sites are classified as follows: metal top site (denoted as empty), metal–metal bridge site (B), metal–nitrogen bridge site (^), carbon site (C), and nitrogen site (N). Volcano plots of the exchange current density ($i_0$) as a function of $\Delta G_{H*}$ for (b) $TM_2$@0N, (c) $TM_2$@1$N_x$, (d) $TM_2$@2$N_x$, (e) $TM_2$@3$N_x$, (f) $TM_2$@4$N_x$, (g) $TM_2$@5$N_x$.

With the BAM MLP validated, we applied the activity branch of the joint screening framework to evaluate HER performance across all 460 $TM_2$@$N_x$–Gr candidates. The hydrogen adsorption free energy ($\Delta G_{H*}$) was used as the primary activity descriptor, as near-thermoneutral hydrogen binding is prerequisite for optimal HER catalytic performance. As shown in **Figure 5b-g and Figure S2**, the exchange current density ($i_0$) exhibits a well-defined volcano-type dependence on $\Delta G_{H*}$, in accordance with the Sabatier principle[59], with peak activity achieved near $\Delta G_{H*} \approx 0$ eV. Increasing nitrogen coordination induces a systematic modulation of $\Delta G_{H*}$, reflecting progressive tuning of the electronic structure of the diatomic metal centers, and intermediate coordination environments (2N–4N) exhibit the higher density of active sites located near the volcano apex. Representative near-optimal candidates include $Ti_2$@2$N_a$, $Mn_2$@2$N_a$, $Fe_2$@2$N_a$, $Cu_2$@2$N_a$, $Rh_2$@2$N_a$, $Zr_2$@2$N_a$, $Zr_2$@2$N_b$, $Zr_2$@2$N_c$, $Nb_2$@2$N_c$, $Zr_2$@2$N_d$, $Mn_2$@2$N_e$, $Mn_2$@2$N_f$, $Ti_2$@3$N_a$, $Au_2$@3$N_a$, $Fe_2$@3$N_a$, $Pd_2$@3$N_b$, $Rh_2$@3$N_c$, $Rh_2$@3$N_d$, $Au_2$@3$N_d$, $V_2$@4$N_a$, $Ti_2$@4$N_b$, $Pd_2$@4$N_b$, $Ti_2$@4$N_c$, $Cr_2$@4$N_d$, $Ni_2$@4$N_d$, $Cu_2$@4$N_d$.

To further identify active HER catalysts, a threshold criterion of $|\Delta G_{H*}| < 0.10$ eV was adopted as the criterion for near-optimal hydrogen adsorption, yielding 26 shortlisted candidates for subsequent electronic feasibility screening and carbon-cost evaluation. The corresponding $\Delta G_{H*}$ values are provided in **Figures S3–S6** for quantitative comparison across the full set of coordination environments. The global activity trends are further summarized in **Figure 5a**, which correlates the HER overpotential ($\eta_{HER}$, upper triangle) with the most stable hydrogen adsorption configurations (H*, lower triangle). Systems with low HER overpotentials ($\eta_{HER}$) predominantly favor hydrogen adsorption at metal-centered sites, including metal top (empty) and metal–metal bridge (B) configurations. In contrast, adsorption at heteroatom sites (C and N) generally exhibits less favorable energetics, resulting in higher HER overpotentials. The prevalence of metal–nitrogen bridge sites (^) increases with nitrogen coordination, highlighting the influence of local coordination geometry on hydrogen adsorption and HER activity. The 26 shortlisted candidates encompass diverse metals, including earth-abundant Mn, Fe, and Cu as well as precious Rh, Au, and Pd, underscoring the limitations of activity-based screening alone in distinguishing promising catalysts. This convergence exposes a critical blind spot of activity-only screening: as will be demonstrated in Section 2.6, these candidates span nearly five orders of magnitude in carbon social cost, making carbon-cost evaluation not a refinement of the screening workflow, but an indispensable component of it.

### 2.5 Electronic Structure Screening and Charge Transport Feasibility

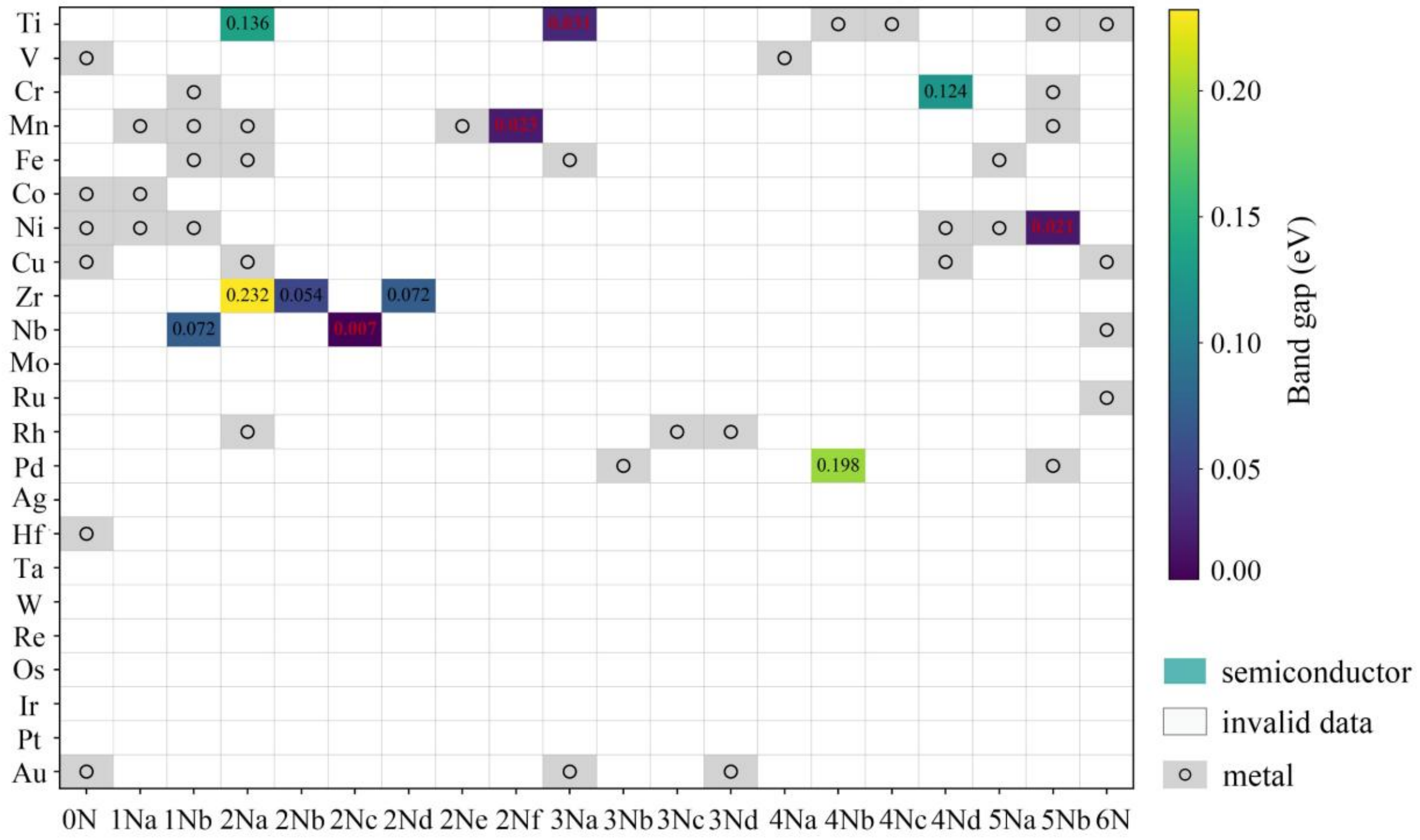

**Figure 6.** Heatmap of calculated band gaps for TM2@Nx-Gr systems (TM = Ti ~ Au) as a function of nitrogen coordination number (x = 0–6). Colored squares represent semiconducting systems with band gap values (in eV) indicated, while open circles denote metallic behavior. White squares indicate invalid data.

Having screened candidates by thermodynamic stability and HER activity, band-gap analysis serves as the final physical screening filter before carbon-cost evaluation. Efficient electrical conductivity is a prerequisite for high HER performance: rapid electron transport to the active sites is essential for fast proton–electron coupling, and poor conductivity introduces interfacial charge-transfer resistance that limits catalytic kinetics even when hydrogen adsorption is thermodynamically favorable. As summarized in **Figure 6** and **Figure S7-13**, the majority of 26 shortlisted candidates exhibit metallic behavior (40 configurations across the full library identified as metals, marked by open circles), with continuous electronic states across the Fermi level enabling rapid electron-transfer kinetics. The remaining semiconducting configurations, including $Ti_2@2N_a$, $Nb_2@1N_b$, $Zr_2@2N_a$, $Zr_2@2N_b$, $Nb_2@2N_c$, $Zr_2@2N_d$, $Mn_2@2N_f$, $Ti_2@3N_a$, $Pd_2@4N_b$, $Cr_2@4N_d$, and $Ni_2@5N_b$, all exhibit narrow band gaps below 0.25 eV; extreme cases such as $Nb_2@2N_c$ (0.007 eV) and $Ni_2@5N_b$ (0.021 eV) approaching metallic behavior and retain sufficient conductivity under practical operating conditions. Formation energy, $\Delta G_{H*}$, and band gap together constitute the complete physical screening branch of the joint framework. The candidates satisfying all three screening criteria exhibit favorable thermodynamic stability, HER activity, and electronic properties. However, their carbon social costs vary by nearly five orders of magnitude, as discussed in Section 2.6, highlighting the importance of incorporating carbon-cost assessment into the screening framework.

### 2.6 Carbon-Cost Evaluation and Activity–Sustainability Ranking

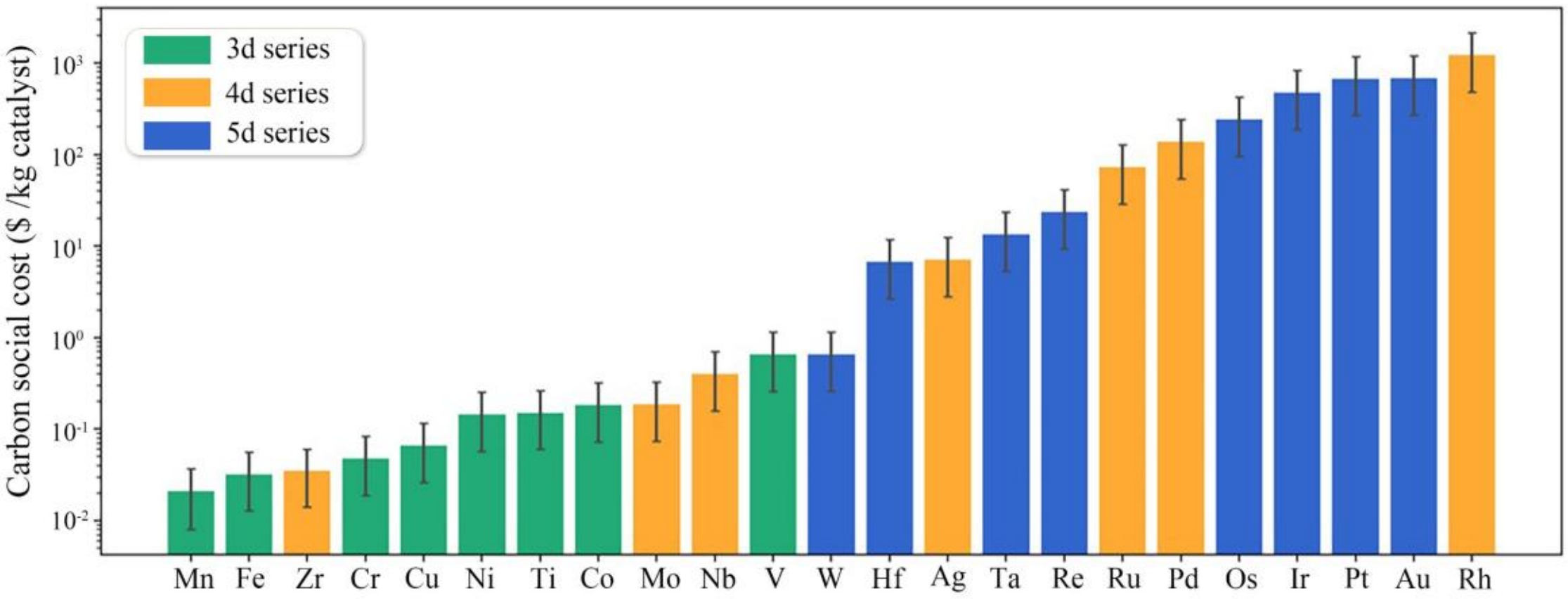

**Figure 7**. Carbon social cost of the $TM_2@N_x$-Gr catalysts. Bars give each metal's embodied $CO_2$ monetized at the central SCC ($141 per tonne $CO_2$); error bars span the low ($56) and high ($248) scenarios. Colours denote the 3d/4d/5d series. On the logarithmic axis, earth-abundant 3d catalysts carry three-to-four orders of magnitude lower carbon social cost than precious-metal analogues.

After the physical screening based on formation energy, $\Delta G_{H*}$, $\eta_{HER}$, and band gap, the same $TM_2@N_x$–Gr catalyst library was evaluated using the carbon-cost layer of the sustainability-aware workflow. This step quantifies the metal-derived embodied $CO_2$ of each catalyst and converts it into carbon social cost, allowing HER activity and upstream metal-related carbon burden to be considered within a single candidate-ranking framework.

Beyond intrinsic activity and electronic structure, the practical promise of non-precious DACs rests on their environmental footprint, which the screening above does not yet quantify. For such evidence to reach policymakers, this footprint is most useful when expressed as a monetary cost rather than as tonnes of $CO_2$ alone, since options across technologies are ultimately weighed in economic terms. To place the 460 $TM_2@N_x$-Gr candidates on a common sustainability scale, we evaluate the carbon social cost embedded in each catalyst, defined as its cradle-to-gate embodied $CO_2$ monetized at the social cost of carbon (SCC): carbon social cost = (metal mass fraction) × (metal embodied GWP) × SCC. Cradle-to-gate global-warming potentials of the constituent metals are taken from the life-cycle synthesis of Nuss and Eckelman[26], and the SCC, the monetary value of one tonne of $CO_2$, is obtained from a machine-learning meta-model. Following the evidence-synthesis approach of Moore et al.[48], we train a random forest on 11,782 published SCC estimates[54] and query it under an expert-elicited distribution of the pure rate of time preference (Drupp et al.)[60] together with persistent (growth-based) climate damages (Methods). This yields a central SCC of 141 USD per tonne $CO_2$ (5-95% range 56 to 248), in line with recent assessments (U.S. EPA 2023[47], 190; Moore et al. 2024[48], 283 USD per tonne $CO_2$). The increase from a literature-level central of about 70 to 141 USD per tonne $CO_2$ reflects the two expert corrections, a lower pure rate of time preference and persistent damages, rather than a simple average of past estimates. The 5th-percentile, mean and 95th-percentile values are carried forward as the low, central and high scenarios to bracket the deep uncertainty intrinsic to the SCC.

The resulting carbon social cost spans nearly five orders of magnitude across the library (**Figure 7**, **Table 1**). At the central SCC (141 USD per tonne $CO_2$), the earth-abundant 3d candidates carry the lowest burden, Mn2 (0.02), Fe2 (0.03), Cr2 (0.05), Cu2 (0.07), Ni2 (0.14) and Ti2 (0.15 USD per kg), together with the abundant Zr2, Mo2, Nb2 and W2 (all below 1 USD per kg). In stark contrast, the noble-metal catalysts incur 138 (Pd2), 674 (Pt2), 678 (Au2) and 1,218 USD per kg (Rh2), the last exceeding the best 3d candidate by a factor of about $5.8 \times 10^4$. The ordering is preserved across the entire SCC range and is therefore robust to the discount rate. Crucially, several of the most economical candidates, $Fe_2@2N_a$, $Mn_2@2N_a$, $Cu_2@4N_d$ and $Ti_2@4N_c$, also sit at the volcano apex (**Figure 5**), showing that low embedded-carbon cost and near-optimal $\Delta G_{H*}$ are simultaneously attainable. These results quantify, in monetary terms, the sustainability rationale for non-precious DAC design and motivate the integrated activity–sustainability ranking shown in **Figure 8**.

**Table 1.** Carbon social cost of the 26 standout $TM_2@N_x$-Gr candidates under the low ($56), central ($141) and high ($248) SCC scenarios (machine-learning meta-model, this work; 5th-percentile, mean and 95th-percentile), ranked by the central value. Embodied $CO_2$ is cradle-to-gate (Nuss & Eckelman 2014); values are a lower bound (metal contribution only), per kg of catalyst.

| Catalyst | Embodied $CO_2$ (kg $CO_2$e/kg) | CSC $56 ($/kg) | CSC $141 ($/kg) | CSC $248 ($/kg) |
|---|---|---|---|---|
| $Mn_2@2N_a$ | 0.15 | 0.008 | 0.021 | 0.037 |
| $Mn_2@2N_e$ | 0.15 | 0.008 | 0.021 | 0.037 |
| $Mn_2@2N_f$ | 0.15 | 0.008 | 0.021 | 0.037 |
| $Fe_2@3N_a$ | 0.23 | 0.013 | 0.032 | 0.056 |
| $Fe_2@2N_a$ | 0.23 | 0.013 | 0.032 | 0.056 |
| $Zr_2@2N_a$ | 0.25 | 0.014 | 0.035 | 0.061 |
| $Zr_2@2N_b$ | 0.25 | 0.014 | 0.035 | 0.061 |
| $Zr_2@2N_c$ | 0.25 | 0.014 | 0.035 | 0.061 |
| $Zr_2@2N_d$ | 0.25 | 0.014 | 0.035 | 0.061 |
| $Cr_2@4N_d$ | 0.34 | 0.019 | 0.048 | 0.084 |
| $Cu_2@4N_d$ | 0.47 | 0.026 | 0.066 | 0.116 |
| $Cu_2@2N_a$ | 0.47 | 0.026 | 0.066 | 0.117 |
| $Ni_2@4N_d$ | 1.02 | 0.057 | 0.143 | 0.252 |
| $Ti_2@4N_b$ | 1.06 | 0.060 | 0.150 | 0.264 |
| $Ti_2@4N_c$ | 1.06 | 0.060 | 0.150 | 0.264 |
| $Ti_2@3N_a$ | 1.07 | 0.060 | 0.151 | 0.265 |
| $Ti_2@2N_a$ | 1.07 | 0.060 | 0.151 | 0.266 |
| $Nb_2@2N_c$ | 2.85 | 0.160 | 0.402 | 0.707 |
| $V_2@4N_a$ | 4.59 | 0.257 | 0.647 | 1.139 |
| $Pd_2@4N_b$ | 976.84 | 54.703 | 137.735 | 242.257 |
| $Pd_2@3N_b$ | 979.15 | 54.833 | 138.061 | 242.830 |
| $Au_2@3N_a$ | 4,806.45 | 269.161 | 677.709 | 1,192 |
| $Au_2@3N_d$ | 4,806.45 | 269.161 | 677.709 | 1,192 |
| $Rh_2@3N_c$ | 8,637.31 | 483.689 | 1,218 | 2,142 |
| $Rh_2@3N_d$ | 8,637.31 | 483.689 | 1,218 | 2,142 |
| $Rh_2@2N_a$ | 8,657.97 | 484.846 | 1,221 | 2,147 |
| **Pt/C 5 wt% (benchmark)** | **625** | **35.0** | **88.1** | **155.0** |

*Carbon social cost values are a lower bound (the carbon-free N-graphene scaffold and synthesis steps are excluded) and are expressed per kilogram of catalyst.*

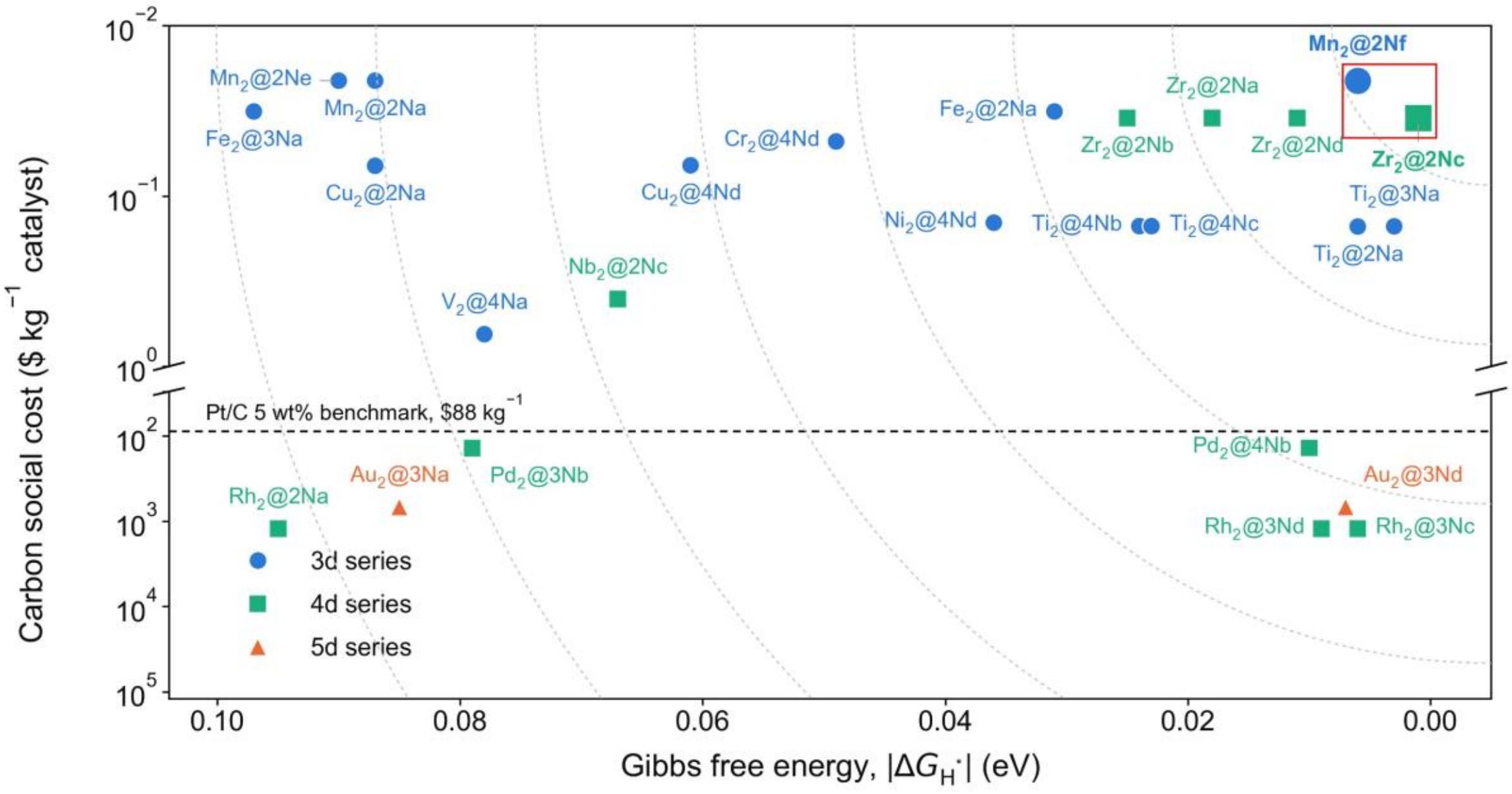

**Figure 8.** Pareto frontier of HER activity versus carbon social cost for the 26 shortlisted $TM_2@N_x$–Gr candidates, blue circles, green squares, and orange triangles denote the 3d, 4d, and 5d series, respectively. The dashed line marks the Pt/C (5 wt%) benchmark (88 USD per kg).

**Figure 8** integrates ML-predicted HER activity with SCC-based carbon-cost assessment to establish a unified framework for evaluating both catalytic performance and environmental sustainability. For the 26 shortlisted $TM_2@N_x$–Gr candidates, $|\Delta G_{H^*}|$ is plotted against the metal-derived carbon social cost calculated at the central SCC value of 141 USD per tonne $CO_2$. The results reveal that catalysts with comparable HER activity can exhibit carbon costs differing by several orders of magnitude, highlighting the limitations of conventional activity-only screening. Notably, several Mn-, Fe-, Zr-, and Ti-based DACs achieve $|\Delta G_{H*}| < 0.10$ eV while exhibiting substantially lower carbon costs than their precious-metal counterparts. By jointly considering these two descriptors, the ML–SCC framework identifies $Mn_2$@$2N_f$ and $Zr_2$@$2N_c$ as Pareto-optimal candidates, offering distinct trade-offs between catalytic activity and carbon cost. Since SCC acts as a common positive scaling factor, their Pareto-optimal status remains unchanged across different SCC scenarios, demonstrating the robustness of the screening results. More broadly, this integrated approach extends computational catalyst screening beyond activity-driven optimization by incorporating environmental costs directly into candidate selection, providing a generalizable strategy for discovering high-performance and sustainable electrocatalysts.

## 3. Conclusion

We introduced a unified machine learning framework that integrates the social cost of carbon as a co-equal optimization criterion alongside catalytic activity in a single catalyst discovery pipeline. A BAM machine learning potential screens 460 $TM_2$@$N_x$–Gr dual-atom catalysts for binding energy, stability, HER activity at DFT accuracy, while a random forest SCC meta-model simultaneously monetizes the metal-derived embodied $CO_2$ of each candidate. The 26 shortlisted candidates exhibit comparable HER activity but differ in carbon social cost by a factor of approximately 58,000, highlighting the importance of integrating environmental considerations into catalyst screening. Pareto analysis identifies $Mn_2$@2Nf and $Zr_2$@2Nc as promising candidates, offering HER activity comparable to that of precious metal catalysts at substantially lower carbon social costs. These results demonstrate that the social cost of carbon can serve as a rigorous quantitative criterion in computational material discovery, and establish sustainability-aware ML screening as an effective strategy for redirecting catalyst design toward earth-abundant, environmentally responsible alternatives without compromise in performance.

## 4. Computational Methods

Spin-polarized DFT calculations were performed using the Vienna Ab initio Simulation Package (VASP)[61-63] to optimize the structures and evaluate their electronic properties. A plane-wave cutoff energy of 500 eV was employed, and the energy convergence criterion was set to $10^{-5}$ eV. For the 2D systems considered, the revised Perdew–Burke–Ernzerhof (RPBE)[64]

exchange–correlation functional, optimized for surface systems, was used in conjunction with the Tkatchenko–Scheffler (TS) dispersion correction.[65] The Hubbard $U_{eff}$ correction was applied to the 3d transition metals, with values of 2.58, 2.72, 2.79, 3.06, 3.29, 3.42, 3.40 and 3.87 eV for Ti through Cu, respectively, as adopted from the literature.[66]As our study primarily focuses on surface phenomena, the RPBE(U)+TS approach is adopted.

### 4.1 Calculating details for HER

To evaluate the Gibbs free energy for the adsorption of H atoms ($\Delta G_{H*}$) for the hydrogen evolution reaction (HER), we employed the computational standard hydrogen electrode (SHE) model proposed by Nørskov et al.[67], in which the chemical potential of a proton–electron pair ($H^+$ + $e^-$) is approximated as one-half of the total energy of an $H_2$ molecule under standard conditions. Under acidic conditions, HER proceeds through the following elementary reaction step:

$$* + H^+ + e^- \rightarrow H* \rightarrow * + 0.5\ H_2 \tag{1}$$

The $\Delta G_{H*}$ was calculated using:

$$\Delta G_{H*} = (E_{H*} - E_{*} - 0.5E_{H2}) + (ZPE_{H*} - ZPE_{*} - 0.5ZPE_{H2}) - T(S_{H*} - S_{*} - 0.5S_{H2}) \tag{2}$$

The formation energies are evaluated according to the following equation:

$$E_f = E_{TM2@Nx\text{–}Gr} - E_{xN\text{–}Gr} - 2(E_{TM\text{-}bulk}/n) \tag{3}$$

$E_{TM2@Nx\text{-}Gr}$ is the total energy of the $TM_2@N_x$-Gr. $E_{xN\text{–}Gr}$ represents the total energy of the nitrogen-coordinated graphene substrate without two metal atoms. $E_{TM\text{-}bulk}$ is the total energy of the bulk transition metal, and n is the number of atoms in the corresponding number of metal atoms in the metal bulk.

### 4.2 Volcano Curve

In the volcano plot, the hydrogen adsorption Gibbs free energy ($\Delta G_{H*}$) serves as a key catalytic descriptor to describe the variation in the exchange current density ($i_o$) for HER, following Nørskov's model.[9] The exchange current density at pH = 0 is calculated as follows:

$$i_0 = -ek_0\left[1 + \exp\left(\frac{|\Delta G_{H^*}|}{k_B T}\right)\right]^{-1} \tag{4}$$

where $k_0$ and $k_B$ denote the rate constant and the Boltzmann constant, respectively, and $k_0$ is set to 1.

### 4.3 Machine learning Method

A machine learning interatomic potential was trained on a DFT dataset comprising 26,406 unique configurations using the equivariant graph neural network RACE as implemented in BAM-torch[58]. Atomic environments were encoded as graphs with a 6.0 Å neighbour cutoff, a hidden representation of $128\times0e + 128\times1o + 128\times2e$, a maximum angular momentum of $l$max = 2 and three interaction layers; atomic forces were obtained as analytical gradients of the predicted energy. Training was carried out on NVIDIA A100 80 GB GPU hardware using the Adam optimizer and minimized a weighted Huber loss[68] of energy and force residuals, $\lambda_E:\lambda_F$ = 30:1 and $\delta = 0.1$. The learning rate was initialized at $1 \times 10^{-3}$ and reduced by a factor of 0.5 on validation-loss plateaus until convergence after approximately 590 epochs. Full hyperparameters are provided in **Table S2**.

### 4.4 Social cost of carbon

The social cost of carbon (SCC) was derived with a machine-learning meta-model, following the evidence-synthesis approach of Moore et al.[48] A random forest (500 trees) was trained to predict log10(SCC, USD per tonne $CO_2$) from the study assumptions, the pure rate of time preference, publication year, elasticity of intertemporal substitution, risk and equity treatment, and damage-function form, using the meta-analytic database of published SCC estimates compiled by Tol[54] (the 11,782 estimates with a positive value and a reported pure rate of time preference), with each estimate weighted by the product of its quality and censoring weights. Five-fold cross-validation gave a median multiplicative error of 1.9x, consistent with the order-of-magnitude dispersion intrinsic to SCC estimates; the pure rate of time preference (feature importance 0.33) and publication year (0.23) were the dominant determinants. A synthetic SCC distribution was then generated by querying the trained model over recent (2016 onward) modelling assumptions while drawing the pure rate of time preference from the expert-elicited distribution of Drupp et al.[60] (median 0.5%, mean 1.1%) and imposing persistent (growth-based) climate damages, mirroring the two expert corrections identified by Moore et al. This yields a central (mean) SCC of 141 USD per tonne $CO_2$ with a 5-95% range of 56 to 248, bracketing the U.S. EPA (2023) central value of 190; the 5th-percentile, mean and 95th-percentile values are carried forward as the low, central and high scenarios. Each catalyst's carbon social cost is the product of its cradle-to-gate embodied $CO_2$, computed from the metal mass fraction and the metal-specific global-warming potentials of Nuss and Eckelman[26], and the SCC. Full results for all 460 catalysts are provided in the Supporting Information (**Table S3**); the model code, data, and parameters are available from the authors on request.

**Supporting Information**

Supporting Information is available from the Wiley Online Library or from the author.

**Acknowledgements**

This work was supported by National Research Foundation of Korea (NRF) funded by the Korean government (Ministry of Science and ICT(MSIT)) (RS-2022-NR072058, RS-2024-00407680, RS-2023-00257666) and Institute for Basic Science (IBS-R036-D1). This work was also supported by the National Research Foundation of Korea (NRF) grant funded by the Korea government (MSIT and MOE) (RS-2026-25501632). YZ, HJK, and CWM are grateful for the computational support from the Korea Institute of Science and Technology Information (KISTI) for the Nurion cluster (KSC-2024-CRE-0358, KSC-2024-CRE-0356, KSC-2025-CRE-0093). Computational work for this research was partially performed on the Olaf supercomputer supported by IBS Research Solution Center on the GPU cluster supported by NIPA. S.Y. acknowledges support from JSPS KAKENHI Grant Numbers JP23K17086, JP23K18816, and JP25K16629.

**Conflict of Interests**

The authors declare no conflict of interests.

Supporting Information

# Sustainability-Aware Discovery of Dual-Atom Hydrogen Evolution Catalysts via a Unified Machine Learning Framework for Activity and Carbon Cost

*Yanmei Zang[1†*], Hyun Gyu Park[1†], Gi Beom Sim[1], Tae Hyeon Park[1], Ho Jin Lee[1], Xiaorong Zou[2], D. ChangMo Yang[1], Soohaeng Yoo Willow[1], Hye Jung Kim[2*], Sunbin Yoo[1,3,4*] Chang Woo Myung[1,2,3,4*]*

[1]Department of Energy Science, Sungkyunkwan University, 2066 Seobu-ro, Suwon 16419, Republic of Korea

[2]Center for 2D Quantum Heterostructures (2DQH), Institute for Basic Science (IBS), Suwon 16419, Republic of Korea

[3]Department of Energy, Sungkyunkwan University, Seobu-ro 2066, Suwon 16419, Republic of Korea

[4]SKKU National Lab for Intelligent Energy Solution Technology (SIEST), Sungkyunkwan University, Seobu-ro 2066, Suwon 16419, Republic of Korea

*E-mail: yanmei@skku.edu (Y.Z.); hjkim75@skku.edu (H. J. K.); sunbinyoo@skku.edu (S. Y.); cwmyung@skku.edu (C. W. M.)

†Contributed equally to this work

**Table S1.** Formation energies (eV) of $TM_2@N_x$–Gr catalysts with different coordination environments.

| system | 0N | $1N_a$ | $1N_b$ | $2N_a$ | $2N_b$ | $2N_c$ | $2N_d$ | $2N_e$ | $2N_f$ | $3N_a$ |
|---|---|---|---|---|---|---|---|---|---|---|
| $Ti_2$@ | -6.41 | -6.16 | -7.21 | -6.03 | -6.38 | -7.39 | -8.85 | -5.74 | -5.34 | -6.57 |
| $V_2$@ | -6.87 | -6.63 | -7.62 | -6.50 | -6.73 | -7.74 | -9.18 | -6.15 | -5.72 | -6.80 |
| $Cr_2$@ | -7.55 | -7.49 | -8.61 | -7.36 | -7.74 | -8.61 | -10.32 | -7.08 | -6.71 | -7.94 |
| $Mn_2$@ | -11.65 | -11.27 | -12.19 | -11.14 | -11.15 | -11.77 | -13.25 | -10.70 | -10.31 | -10.81 |
| $Fe_2$@ | -6.70 | -6.37 | -6.98 | -6.24 | -5.89 | -6.40 | -7.95 | -5.51 | -5.29 | -5.55 |
| $Co_2$@ | -7.12 | -6.88 | -7.39 | -6.75 | -6.41 | -6.57 | -7.39 | -6.18 | -5.81 | -5.73 |
| $Ni_2$@ | -7.43 | -7.02 | -7.33 | -6.89 | -6.32 | -6.31 | -7.87 | -6.23 | -5.84 | -5.47 |
| $Cu_2$@ | -5.45 | -4.44 | -4.86 | -4.31 | -3.15 | -3.33 | -5.34 | -3.20 | -2.83 | -2.68 |
| $Zr_2$@ | -6.19 | -6.12 | -7.31 | -5.99 | -6.02 | -7.10 | -8.49 | -5.66 | -5.24 | -6.14 |
| $Nb_2$@ | -5.45 | -5.18 | -6.08 | -5.05 | -4.70 | -5.68 | -6.97 | -4.28 | -4.06 | -4.71 |
| $Mo_2$@ | -4.34 | -4.04 | -4.79 | -3.92 | -3.58 | -4.15 | -5.54 | -3.32 | -2.89 | -3.34 |
| $Ru_2$@ | -5.20 | -4.77 | -5.19 | -4.64 | -4.24 | -4.60 | -6.24 | -3.77 | -3.49 | -3.89 |
| $Rh_2$@ | -5.95 | -5.69 | -6.25 | -5.57 | -5.71 | -5.85 | -7.47 | -5.17 | -4.81 | -5.02 |
| $Pd_2$@ | -6.39 | -5.95 | -6.33 | -5.82 | -5.22 | -5.16 | -6.70 | -5.13 | -4.70 | -4.22 |
| $Ag_2$@ | -3.02 | -1.93 | -2.40 | -1.80 | -0.42 | -0.67 | -2.56 | -0.65 | -0.23 | 0.02 |
| $Hf_2$@ | -5.60 | -5.54 | -6.75 | -5.41 | -5.54 | -6.73 | -8.07 | -5.11 | -4.66 | -5.73 |
| $Ta_2$@ | -4.63 | -4.34 | -5.27 | -4.21 | -3.90 | -4.77 | -6.10 | -3.51 | -3.14 | -3.78 |
| $W_2$@ | -3.52 | -3.19 | -4.07 | -3.06 | -2.79 | -3.46 | -4.81 | -2.48 | -2.11 | -2.63 |
| $Re_2$@ | -4.57 | -4.06 | -4.50 | -3.93 | -3.29 | -3.48 | -5.05 | -3.02 | -2.73 | -2.91 |
| $Os_2$@ | -5.00 | -4.35 | -4.76 | -4.22 | -3.36 | -3.85 | -5.52 | -3.27 | -3.01 | -3.20 |
| $Ir_2$@ | -6.49 | -5.96 | -6.65 | -5.83 | -5.60 | -5.83 | -7.47 | -5.24 | -4.85 | -5.05 |
| $Pt_2$@ | -8.29 | -7.61 | -8.05 | -7.48 | -6.36 | -6.28 | -7.85 | -6.53 | -6.15 | -5.39 |
| $Au_2$@ | -5.50 | -4.17 | -4.61 | -4.04 | -2.29 | -2.40 | -4.42 | -2.51 | -2.13 | -1.90 |

| system | $3N_b$ | $3N_c$ | $3N_d$ | $4N_a$ | $4N_b$ | $4N_c$ | $4N_d$ | $5N_a$ | $5N_b$ | 6N |
|---|---|---|---|---|---|---|---|---|---|---|
| $Ti_2$@ | -6.90 | -8.38 | -6.61 | -8.46 | -7.45 | -9.40 | -10.63 | -8.08 | -8.75 | -9.45 |
| $V_2$@ | -7.25 | -8.73 | -6.95 | -8.84 | -7.92 | -9.84 | -11.14 | -8.58 | -9.19 | -9.95 |
| $Cr_2$@ | -8.26 | -9.59 | -8.09 | -9.91 | -9.18 | -11.05 | -12.15 | -9.59 | -10.66 | -10.96 |
| $Mn_2$@ | -11.67 | -12.75 | -11.01 | -12.98 | -11.61 | -13.40 | -14.53 | -11.97 | -12.69 | -12.69 |
| $Fe_2$@ | -6.41 | -7.38 | -5.71 | -7.92 | -5.99 | -8.34 | -8.08 | -5.52 | -7.28 | -6.71 |
| $Co_2$@ | -6.93 | -7.56 | -5.15 | -8.17 | -6.39 | -8.27 | -8.15 | -5.59 | -7.26 | -6.58 |
| $Ni_2$@ | -6.84 | -7.30 | -5.63 | -8.00 | -6.08 | -7.89 | -8.25 | -5.69 | -7.01 | -6.07 |
| $Cu_2$@ | -3.67 | -4.31 | -3.10 | -4.50 | -3.73 | -5.30 | -6.36 | -3.80 | -4.57 | -4.32 |
| $Zr_2$@ | -6.54 | -8.09 | -6.25 | -8.04 | -7.00 | -9.01 | -10.18 | -7.62 | -8.32 | -8.63 |
| $Nb_2$@ | -5.22 | -6.67 | -4.73 | -6.50 | -5.49 | -7.53 | -8.55 | -5.99 | -6.72 | -6.90 |
| $Mo_2$@ | -4.10 | -5.14 | -3.30 | -5.42 | -4.01 | -5.98 | -6.95 | -4.39 | -5.17 | -5.40 |
| $Ru_2$@ | -4.76 | -5.59 | -4.00 | -6.28 | -4.97 | -6.78 | -7.45 | -4.89 | -6.22 | -5.65 |
| $Rh_2$@ | -6.23 | -6.84 | -5.23 | -7.81 | -6.12 | -7.91 | -8.11 | -5.55 | -7.02 | -5.98 |
| $Pd_2$@ | -5.74 | -6.15 | -4.46 | -6.91 | -4.70 | -6.65 | -6.66 | -4.10 | -5.31 | -4.70 |
| $Ag_2$@ | -0.94 | -1.66 | -0.32 | -1.80 | -0.94 | -2.47 | -3.50 | -0.94 | -1.35 | -0.82 |
| $Hf_2$@ | -6.06 | -7.71 | -5.84 | -7.61 | -6.60 | -8.65 | -9.83 | -7.27 | -7.96 | -8.20 |
| $Ta_2$@ | -4.42 | -5.76 | -3.87 | -5.69 | -4.50 | -6.66 | -7.64 | -5.08 | -5.71 | -5.99 |
| $W_2$@ | -3.31 | -4.44 | -2.57 | -4.63 | -3.40 | -5.37 | -6.33 | -3.77 | -4.56 | -4.80 |
| $Re_2$@ | -3.81 | -4.46 | -2.81 | -5.03 | -3.47 | -5.37 | -6.44 | -3.88 | -4.57 | -4.74 |
| $Os_2$@ | -3.88 | -4.84 | -3.28 | -5.35 | -4.08 | -5.88 | -6.65 | -4.09 | -5.16 | -4.65 |
| $Ir_2$@ | -6.12 | -6.82 | -5.23 | -7.54 | -5.91 | -7.71 | -8.01 | -5.45 | -6.54 | -5.25 |
| $Pt_2$@ | -6.88 | -7.26 | -5.61 | -7.76 | -5.62 | -7.53 | -7.50 | -4.94 | -5.83 | -5.11 |
| $Au_2$@ | -2.81 | -3.38 | -2.18 | -3.95 | -2.73 | -4.15 | -5.18 | -2.62 | -2.92 | -1.67 |

**Table S2.** Hyperparameters used for training the machine learning potential (MLP) model.

| Parameter | Value |
|---|---|
| **Architecture** | RACE (equivariant GNN) |
| **Cutoff radius** | 6.0 Å |
| **Hidden channels** | 128×0$e$ + 128×1$o$ + 128×2$e$ |
| **Feature dimension** | 64 |
| **$l$max** | 2 |
| **Radial basis functions** | 8 |
| **Interaction layers** | 3 |
| **Optimizer** | Adam (default $\beta 1$, $\beta 2$, $\varepsilon$) |
| **Batch size** | 16 |
| **Initial learning rate** | $1.0 \times 10^{-3}$ |
| **Final learning rate** | $3.125 \times 10^{-5}$ |
| **Weight decay** | $5.0 \times 10^{-6}$ |
| **LR scheduler** | ReduceLROnPlateau (factor 0.5, patience 30, threshold 0.01) |
| **Loss** | Huber, $\delta = 0.1$ |
| **Loss weights ($\lambda E, \lambda F, \lambda\sigma$)** | (30, 1, 0) |
| **Train/validation/test split** | 21,528/2,691/2,693 |
| **Data/weight seeds** | 1300/1301 |
| **Hardware** | 1 × NVIDIA A100 80 GB |

**Table S3.** Embodied $CO_2$(cradle-to-gate) and carbon social cost of all 460 $TM_2@N_x$-Gr catalysts under the low ($56), central ($141) and high ($248) SCC scenarios (machine-learning meta-model, this work; 5th-percentile, mean and 95th-percentile), ranked by the central value. Embodied $CO_2$ from metal mass fraction x metal-specific GWP (Nuss & Eckelman 2014); SCC from a random-forest meta-model trained on the Tol database, queried with the expert discount-rate distribution of Drupp et al. (2018), following Moore et al. (2024) (Methods). Lower bound (metal contribution only), per kg of catalyst.

| **Catalyst** | **Embodied $CO_2$ (kg CO2e/kg)** | **CSC $56 ($/kg)** | **CSC $141 ($/kg)** | **CSC $248 ($/kg)** |
|---|---|---|---|---|
| $Mn_2$@6N | 0.147 | 0.0082 | 0.0208 | 0.0365 |
| $Mn_2$@5Na | 0.148 | 0.0083 | 0.0208 | 0.0366 |
| $Mn_2$@5Nb | 0.148 | 0.0083 | 0.0208 | 0.0366 |
| $Mn_2$@4Na | 0.148 | 0.0083 | 0.0209 | 0.0367 |
| $Mn_2$@4Nb | 0.148 | 0.0083 | 0.0209 | 0.0367 |
| $Mn_2$@4Nc | 0.148 | 0.0083 | 0.0209 | 0.0367 |
| $Mn_2$@4Nd | 0.148 | 0.0083 | 0.0209 | 0.0367 |
| $Mn_2$@3Na | 0.148 | 0.0083 | 0.0209 | 0.0368 |
| $Mn_2$@3Nb | 0.148 | 0.0083 | 0.0209 | 0.0368 |
| $Mn_2$@3Nc | 0.148 | 0.0083 | 0.0209 | 0.0368 |
| $Mn_2$@3Nd | 0.148 | 0.0083 | 0.0209 | 0.0368 |
| $Mn_2$@2Na | 0.149 | 0.0083 | 0.0210 | 0.0369 |
| $Mn_2$@2Nb | 0.149 | 0.0083 | 0.0210 | 0.0369 |
| $Mn_2$@2Nc | 0.149 | 0.0083 | 0.0210 | 0.0369 |
| $Mn_2$@2Nd | 0.149 | 0.0083 | 0.0210 | 0.0369 |
| $Mn_2$@2Ne | 0.149 | 0.0083 | 0.0210 | 0.0369 |
| $Mn_2$@2Nf | 0.149 | 0.0083 | 0.0210 | 0.0369 |
| $Mn_2$@1Na | 0.149 | 0.0084 | 0.0210 | 0.0370 |
| $Mn_2$@1Nb | 0.149 | 0.0084 | 0.0210 | 0.0370 |
| $Mn_2$@0N | 0.150 | 0.0084 | 0.0211 | 0.0371 |
| $Fe_2$@6N | 0.224 | 0.0125 | 0.0316 | 0.0555 |
| $Fe_2$@5Na | 0.225 | 0.0126 | 0.0317 | 0.0557 |
| $Fe_2$@5Nb | 0.225 | 0.0126 | 0.0317 | 0.0557 |
| $Fe_2$@4Na | 0.225 | 0.0126 | 0.0317 | 0.0558 |
| $Fe_2$@4Nb | 0.225 | 0.0126 | 0.0317 | 0.0558 |
| $Fe_2$@4Nc | 0.225 | 0.0126 | 0.0317 | 0.0558 |
| $Fe_2$@4Nd | 0.225 | 0.0126 | 0.0317 | 0.0558 |
| $Fe_2$@3Na | 0.226 | 0.0126 | 0.0318 | 0.0560 |
| $Fe_2$@3Nb | 0.226 | 0.0126 | 0.0318 | 0.0560 |
| $Fe_2$@3Nc | 0.226 | 0.0126 | 0.0318 | 0.0560 |
| $Fe_2$@3Nd | 0.226 | 0.0126 | 0.0318 | 0.0560 |
| $Fe_2$@2Na | 0.226 | 0.0127 | 0.0319 | 0.0561 |
| $Fe_2$@2Nb | 0.226 | 0.0127 | 0.0319 | 0.0561 |
| $Fe_2$@2Nc | 0.226 | 0.0127 | 0.0319 | 0.0561 |
| $Fe_2$@2Nd | 0.226 | 0.0127 | 0.0319 | 0.0561 |
| $Fe_2$@2Ne | 0.226 | 0.0127 | 0.0319 | 0.0561 |
| $Fe_2$@2Nf | 0.226 | 0.0127 | 0.0319 | 0.0561 |
| $Fe_2$@1Na | 0.227 | 0.0127 | 0.0320 | 0.0563 |
| $Fe_2$@1Nb | 0.227 | 0.0127 | 0.0320 | 0.0563 |
| $Fe_2$@0N | 0.228 | 0.0127 | 0.0321 | 0.0564 |
| $Zr_2$@6N | 0.245 | 0.0137 | 0.0346 | 0.0608 |
| $Zr_2$@5Na | 0.246 | 0.0138 | 0.0346 | 0.0609 |
| $Zr_2$@5Nb | 0.246 | 0.0138 | 0.0346 | 0.0609 |
| $Zr_2$@4Na | 0.246 | 0.0138 | 0.0347 | 0.0611 |
| $Zr_2$@4Nb | 0.246 | 0.0138 | 0.0347 | 0.0611 |
| $Zr_2$@4Nc | 0.246 | 0.0138 | 0.0347 | 0.0611 |
| $Zr_2$@4Nd | 0.246 | 0.0138 | 0.0347 | 0.0611 |
| $Zr_2$@3Na | 0.247 | 0.0138 | 0.0348 | 0.0612 |
| $Zr_2$@3Nb | 0.247 | 0.0138 | 0.0348 | 0.0612 |
| $Zr_2$@3Nc | 0.247 | 0.0138 | 0.0348 | 0.0612 |
| $Zr_2$@3Nd | 0.247 | 0.0138 | 0.0348 | 0.0612 |
| $Zr_2$@2Na | 0.247 | 0.0139 | 0.0349 | 0.0614 |
| $Zr_2$@2Nb | 0.247 | 0.0139 | 0.0349 | 0.0614 |
| $Zr_2$@2Nc | 0.247 | 0.0139 | 0.0349 | 0.0614 |
| $Zr_2$@2Nd | 0.247 | 0.0139 | 0.0349 | 0.0614 |
| $Zr_2$@2Ne | 0.247 | 0.0139 | 0.0349 | 0.0614 |
| $Zr_2$@2Nf | 0.247 | 0.0139 | 0.0349 | 0.0614 |
| $Zr_2$@1Na | 0.248 | 0.0139 | 0.0350 | 0.0615 |
| $Zr_2$@1Nb | 0.248 | 0.0139 | 0.0350 | 0.0615 |
| $Zr_2$@0N | 0.249 | 0.0139 | 0.0351 | 0.0617 |
| $Cr_2$@6N | 0.337 | 0.0189 | 0.0475 | 0.0836 |
| $Cr_2$@5Na | 0.338 | 0.0189 | 0.0476 | 0.0838 |
| $Cr_2$@5Nb | 0.338 | 0.0189 | 0.0476 | 0.0838 |
| $Cr_2$@4Na | 0.339 | 0.0190 | 0.0478 | 0.0840 |
| $Cr_2$@4Nb | 0.339 | 0.0190 | 0.0478 | 0.0840 |
| $Cr_2$@4Nc | 0.339 | 0.0190 | 0.0478 | 0.0840 |
| $Cr_2$@4Nd | 0.339 | 0.0190 | 0.0478 | 0.0840 |
| $Cr_2$@3Na | 0.340 | 0.0190 | 0.0479 | 0.0843 |
| $Cr_2$@3Nb | 0.340 | 0.0190 | 0.0479 | 0.0843 |
| $Cr_2$@3Nc | 0.340 | 0.0190 | 0.0479 | 0.0843 |
| $Cr_2$@3Nd | 0.340 | 0.0190 | 0.0479 | 0.0843 |
| $Cr_2$@2Na | 0.341 | 0.0191 | 0.0480 | 0.0845 |
| $Cr_2$@2Nb | 0.341 | 0.0191 | 0.0480 | 0.0845 |
| $Cr_2$@2Nc | 0.341 | 0.0191 | 0.0480 | 0.0845 |
| $Cr_2$@2Nd | 0.341 | 0.0191 | 0.0480 | 0.0845 |
| $Cr_2$@2Ne | 0.341 | 0.0191 | 0.0480 | 0.0845 |
| $Cr_2$@2Nf | 0.341 | 0.0191 | 0.0480 | 0.0845 |
| $Cr_2$@1Na | 0.342 | 0.0191 | 0.0482 | 0.0847 |
| $Cr_2$@1Nb | 0.342 | 0.0191 | 0.0482 | 0.0847 |
| $Cr_2$@0N | 0.343 | 0.0192 | 0.0483 | 0.0850 |
| $Cu_2$@6N | 0.466 | 0.0261 | 0.0657 | 0.1156 |
| $Cu_2$@5Na | 0.467 | 0.0262 | 0.0659 | 0.1159 |
| $Cu_2$@5Nb | 0.467 | 0.0262 | 0.0659 | 0.1159 |
| $Cu_2$@4Na | 0.468 | 0.0262 | 0.0661 | 0.1162 |
| $Cu_2$@4Nb | 0.468 | 0.0262 | 0.0661 | 0.1162 |
| $Cu_2$@4Nc | 0.468 | 0.0262 | 0.0661 | 0.1162 |
| $Cu_2$@4Nd | 0.468 | 0.0262 | 0.0661 | 0.1162 |
| $Cu_2$@3Na | 0.470 | 0.0263 | 0.0662 | 0.1165 |
| $Cu_2$@3Nb | 0.470 | 0.0263 | 0.0662 | 0.1165 |
| $Cu_2$@3Nc | 0.470 | 0.0263 | 0.0662 | 0.1165 |
| $Cu_2$@3Nd | 0.470 | 0.0263 | 0.0662 | 0.1165 |
| $Cu_2$@2Na | 0.471 | 0.0264 | 0.0664 | 0.1168 |
| $Cu_2$@2Nb | 0.471 | 0.0264 | 0.0664 | 0.1168 |
| $Cu_2$@2Nc | 0.471 | 0.0264 | 0.0664 | 0.1168 |
| $Cu_2$@2Nd | 0.471 | 0.0264 | 0.0664 | 0.1168 |
| $Cu_2$@2Ne | 0.471 | 0.0264 | 0.0664 | 0.1168 |
| $Cu_2$@2Nf | 0.471 | 0.0264 | 0.0664 | 0.1168 |
| $Cu_2$@1Na | 0.472 | 0.0264 | 0.0666 | 0.1171 |
| $Cu_2$@1Nb | 0.472 | 0.0264 | 0.0666 | 0.1171 |
| $Cu_2$@0N | 0.473 | 0.0265 | 0.0668 | 0.1174 |
| $Ni_2$@6N | 1.012 | 0.0567 | 0.1427 | 0.2510 |
| $Ni_2$@5Na | 1.015 | 0.0568 | 0.1431 | 0.2517 |
| $Ni_2$@5Nb | 1.015 | 0.0568 | 0.1431 | 0.2517 |
| $Ni_2$@4Na | 1.017 | 0.0570 | 0.1435 | 0.2523 |
| $Ni_2$@4Nb | 1.017 | 0.0570 | 0.1435 | 0.2523 |
| $Ni_2$@4Nc | 1.017 | 0.0570 | 0.1435 | 0.2523 |
| $Ni_2$@4Nd | 1.017 | 0.0570 | 0.1435 | 0.2523 |
| $Ni_2$@3Na | 1.020 | 0.0571 | 0.1438 | 0.2530 |
| $Ni_2$@3Nb | 1.020 | 0.0571 | 0.1438 | 0.2530 |
| $Ni_2$@3Nc | 1.020 | 0.0571 | 0.1438 | 0.2530 |
| $Ni_2$@3Nd | 1.020 | 0.0571 | 0.1438 | 0.2530 |
| $Ni_2$@2Na | 1.023 | 0.0573 | 0.1442 | 0.2537 |
| $Ni_2$@2Nb | 1.023 | 0.0573 | 0.1442 | 0.2537 |
| $Ni_2$@2Nc | 1.023 | 0.0573 | 0.1442 | 0.2537 |
| $Ni_2$@2Nd | 1.023 | 0.0573 | 0.1442 | 0.2537 |
| $Ni_2$@2Ne | 1.023 | 0.0573 | 0.1442 | 0.2537 |
| $Ni_2$@2Nf | 1.023 | 0.0573 | 0.1442 | 0.2537 |
| $Ni_2$@1Na | 1.026 | 0.0574 | 0.1446 | 0.2544 |
| $Ni_2$@1Nb | 1.026 | 0.0574 | 0.1446 | 0.2544 |
| $Ni_2$@0N | 1.028 | 0.0576 | 0.1450 | 0.2550 |
| $Ti_2$@6N | 1.059 | 0.0593 | 0.1493 | 0.2626 |
| $Ti_2$@5Na | 1.062 | 0.0595 | 0.1497 | 0.2633 |
| $Ti_2$@5Nb | 1.062 | 0.0595 | 0.1497 | 0.2633 |
| $Ti_2$@4Na | 1.065 | 0.0596 | 0.1501 | 0.2641 |
| $Ti_2$@4Nb | 1.065 | 0.0596 | 0.1501 | 0.2641 |
| $Ti_2$@4Nc | 1.065 | 0.0596 | 0.1501 | 0.2641 |
| $Ti_2$@4Nd | 1.065 | 0.0596 | 0.1501 | 0.2641 |
| $Ti_2$@3Na | 1.068 | 0.0598 | 0.1505 | 0.2648 |
| $Ti_2$@3Nb | 1.068 | 0.0598 | 0.1505 | 0.2648 |
| $Ti_2$@3Nc | 1.068 | 0.0598 | 0.1505 | 0.2648 |
| $Ti_2$@3Nd | 1.068 | 0.0598 | 0.1505 | 0.2648 |
| $Ti_2$@2Na | 1.071 | 0.0600 | 0.1510 | 0.2655 |
| $Ti_2$@2Nb | 1.071 | 0.0600 | 0.1510 | 0.2655 |
| $Ti_2$@2Nc | 1.071 | 0.0600 | 0.1510 | 0.2655 |
| $Ti_2$@2Nd | 1.071 | 0.0600 | 0.1510 | 0.2655 |
| $Ti_2$@2Ne | 1.071 | 0.0600 | 0.1510 | 0.2655 |
| $Ti_2$@2Nf | 1.071 | 0.0600 | 0.1510 | 0.2655 |
| $Ti_2$@1Na | 1.074 | 0.0601 | 0.1514 | 0.2662 |
| $Ti_2$@1Nb | 1.074 | 0.0601 | 0.1514 | 0.2662 |
| $Ti_2$@0N | 1.077 | 0.0603 | 0.1518 | 0.2670 |
| $Co_2$@6N | 1.297 | 0.0726 | 0.1828 | 0.3216 |
| $Co_2$@5Na | 1.300 | 0.0728 | 0.1833 | 0.3224 |
| $Co_2$@5Nb | 1.300 | 0.0728 | 0.1833 | 0.3224 |
| $Co_2$@4Na | 1.304 | 0.0730 | 0.1838 | 0.3233 |
| $Co_2$@4Nb | 1.304 | 0.0730 | 0.1838 | 0.3233 |
| $Co_2$@4Nc | 1.304 | 0.0730 | 0.1838 | 0.3233 |
| $Co_2$@4Nd | 1.304 | 0.0730 | 0.1838 | 0.3233 |
| $Co_2$@3Na | 1.307 | 0.0732 | 0.1843 | 0.3242 |
| $Co_2$@3Nb | 1.307 | 0.0732 | 0.1843 | 0.3242 |
| $Co_2$@3Nc | 1.307 | 0.0732 | 0.1843 | 0.3242 |
| $Co_2$@3Nd | 1.307 | 0.0732 | 0.1843 | 0.3242 |
| $Co_2$@2Na | 1.311 | 0.0734 | 0.1848 | 0.3250 |
| $Co_2$@2Nb | 1.311 | 0.0734 | 0.1848 | 0.3250 |
| $Co_2$@2Nc | 1.311 | 0.0734 | 0.1848 | 0.3250 |
| $Co_2$@2Nd | 1.311 | 0.0734 | 0.1848 | 0.3250 |
| $Co_2$@2Ne | 1.311 | 0.0734 | 0.1848 | 0.3250 |
| $Co_2$@2Nf | 1.311 | 0.0734 | 0.1848 | 0.3250 |
| $Co_2$@1Na | 1.314 | 0.0736 | 0.1853 | 0.3259 |
| $Co_2$@1Nb | 1.314 | 0.0736 | 0.1853 | 0.3259 |
| $Co_2$@0N | 1.318 | 0.0738 | 0.1858 | 0.3268 |
| $Mo_2$@6N | 1.320 | 0.0739 | 0.1862 | 0.3274 |
| $Mo_2$@5Na | 1.324 | 0.0741 | 0.1866 | 0.3282 |
| $Mo_2$@5Nb | 1.324 | 0.0741 | 0.1866 | 0.3282 |
| $Mo_2$@4Na | 1.327 | 0.0743 | 0.1871 | 0.3290 |
| $Mo_2$@4Nb | 1.327 | 0.0743 | 0.1871 | 0.3290 |
| $Mo_2$@4Nc | 1.327 | 0.0743 | 0.1871 | 0.3290 |
| $Mo_2$@4Nd | 1.327 | 0.0743 | 0.1871 | 0.3290 |
| $Mo_2$@3Na | 1.330 | 0.0745 | 0.1875 | 0.3298 |
| $Mo_2$@3Nb | 1.330 | 0.0745 | 0.1875 | 0.3298 |
| $Mo_2$@3Nc | 1.330 | 0.0745 | 0.1875 | 0.3298 |
| $Mo_2$@3Nd | 1.330 | 0.0745 | 0.1875 | 0.3298 |
| $Mo_2$@2Na | 1.333 | 0.0747 | 0.1880 | 0.3306 |
| $Mo_2$@2Nb | 1.333 | 0.0747 | 0.1880 | 0.3306 |
| $Mo_2$@2Nc | 1.333 | 0.0747 | 0.1880 | 0.3306 |
| $Mo_2$@2Nd | 1.333 | 0.0747 | 0.1880 | 0.3306 |
| $Mo_2$@2Ne | 1.333 | 0.0747 | 0.1880 | 0.3306 |
| $Mo_2$@2Nf | 1.333 | 0.0747 | 0.1880 | 0.3306 |
| $Mo_2$@1Na | 1.336 | 0.0748 | 0.1884 | 0.3314 |
| $Mo_2$@1Nb | 1.336 | 0.0748 | 0.1884 | 0.3314 |
| $Mo_2$@0N | 1.340 | 0.0750 | 0.1889 | 0.3322 |
| $Nb_2$@6N | 2.824 | 0.1582 | 0.3982 | 0.7004 |
| $Nb_2$@5Na | 2.831 | 0.1585 | 0.3992 | 0.7021 |
| $Nb_2$@5Nb | 2.831 | 0.1585 | 0.3992 | 0.7021 |
| $Nb_2$@4Na | 2.838 | 0.1589 | 0.4002 | 0.7039 |
| $Nb_2$@4Nb | 2.838 | 0.1589 | 0.4002 | 0.7039 |
| $Nb_2$@4Nc | 2.838 | 0.1589 | 0.4002 | 0.7039 |
| $Nb_2$@4Nd | 2.838 | 0.1589 | 0.4002 | 0.7039 |
| $Nb_2$@3Na | 2.845 | 0.1593 | 0.4012 | 0.7056 |
| $Nb_2$@3Nb | 2.845 | 0.1593 | 0.4012 | 0.7056 |
| $Nb_2$@3Nc | 2.845 | 0.1593 | 0.4012 | 0.7056 |
| $Nb_2$@3Nd | 2.845 | 0.1593 | 0.4012 | 0.7056 |
| $Nb_2$@2Na | 2.852 | 0.1597 | 0.4021 | 0.7073 |
| $Nb_2$@2Nb | 2.852 | 0.1597 | 0.4021 | 0.7073 |
| $Nb_2$@2Nc | 2.852 | 0.1597 | 0.4021 | 0.7073 |
| $Nb_2$@2Nd | 2.852 | 0.1597 | 0.4021 | 0.7073 |
| $Nb_2$@2Ne | 2.852 | 0.1597 | 0.4021 | 0.7073 |
| $Nb_2$@2Nf | 2.852 | 0.1597 | 0.4021 | 0.7073 |
| $Nb_2$@1Na | 2.859 | 0.1601 | 0.4031 | 0.7090 |
| $Nb_2$@1Nb | 2.859 | 0.1601 | 0.4031 | 0.7090 |
| $Nb_2$@0N | 2.866 | 0.1605 | 0.4041 | 0.7108 |
| $V_2$@6N | 4.567 | 0.2557 | 0.6439 | 1.1326 |
| $V_2$@5Na | 4.579 | 0.2564 | 0.6457 | 1.1357 |
| $V_2$@5Nb | 4.579 | 0.2564 | 0.6457 | 1.1357 |
| $V_2$@4Na | 4.592 | 0.2571 | 0.6474 | 1.1388 |
| $V_2$@4Nb | 4.592 | 0.2571 | 0.6474 | 1.1388 |
| $V_2$@4Nc | 4.592 | 0.2571 | 0.6474 | 1.1388 |
| $V_2$@4Nd | 4.592 | 0.2571 | 0.6474 | 1.1388 |
| $V_2$@3Na | 4.604 | 0.2578 | 0.6492 | 1.1419 |
| $V_2$@3Nb | 4.604 | 0.2578 | 0.6492 | 1.1419 |
| $V_2$@3Nc | 4.604 | 0.2578 | 0.6492 | 1.1419 |
| $V_2$@3Nd | 4.604 | 0.2578 | 0.6492 | 1.1419 |
| $W_2$@6N | 4.613 | 0.2583 | 0.6505 | 1.1441 |
| $V_2$@2Na | 4.617 | 0.2585 | 0.6510 | 1.1450 |
| $V_2$@2Nb | 4.617 | 0.2585 | 0.6510 | 1.1450 |
| $V_2$@2Nc | 4.617 | 0.2585 | 0.6510 | 1.1450 |
| $V_2$@2Nd | 4.617 | 0.2585 | 0.6510 | 1.1450 |
| $V_2$@2Ne | 4.617 | 0.2585 | 0.6510 | 1.1450 |
| $V_2$@2Nf | 4.617 | 0.2585 | 0.6510 | 1.1450 |
| $W_2$@5Na | 4.622 | 0.2589 | 0.6518 | 1.1464 |
| $W_2$@5Nb | 4.622 | 0.2589 | 0.6518 | 1.1464 |
| $V_2$@1Na | 4.629 | 0.2593 | 0.6528 | 1.1481 |
| $V_2$@1Nb | 4.629 | 0.2593 | 0.6528 | 1.1481 |
| $W_2$@4Na | 4.632 | 0.2594 | 0.6531 | 1.1487 |
| $W_2$@4Nb | 4.632 | 0.2594 | 0.6531 | 1.1487 |
| $W_2$@4Nc | 4.632 | 0.2594 | 0.6531 | 1.1487 |
| $W_2$@4Nd | 4.632 | 0.2594 | 0.6531 | 1.1487 |
| $W_2$@3Na | 4.641 | 0.2599 | 0.6544 | 1.1510 |
| $W_2$@3Nb | 4.641 | 0.2599 | 0.6544 | 1.1510 |
| $W_2$@3Nc | 4.641 | 0.2599 | 0.6544 | 1.1510 |
| $W_2$@3Nd | 4.641 | 0.2599 | 0.6544 | 1.1510 |
| $V_2$@0N | 4.642 | 0.2600 | 0.6546 | 1.1513 |
| $W_2$@2Na | 4.650 | 0.2604 | 0.6557 | 1.1533 |
| $W_2$@2Nb | 4.650 | 0.2604 | 0.6557 | 1.1533 |
| $W_2$@2Nc | 4.650 | 0.2604 | 0.6557 | 1.1533 |
| $W_2$@2Nd | 4.650 | 0.2604 | 0.6557 | 1.1533 |
| $W_2$@2Ne | 4.650 | 0.2604 | 0.6557 | 1.1533 |
| $W_2$@2Nf | 4.650 | 0.2604 | 0.6557 | 1.1533 |
| $W_2$@1Na | 4.660 | 0.2609 | 0.6570 | 1.1556 |
| $W_2$@1Nb | 4.660 | 0.2609 | 0.6570 | 1.1556 |
| $W_2$@0N | 4.669 | 0.2615 | 0.6583 | 1.1579 |
| $Hf_2$@6N | 47.069 | 2.6359 | 6.6367 | 11.6731 |
| $Hf_2$@5Na | 47.164 | 2.6412 | 6.6501 | 11.6966 |
| $Hf_2$@5Nb | 47.164 | 2.6412 | 6.6501 | 11.6966 |
| $Hf_2$@4Na | 47.259 | 2.6465 | 6.6635 | 11.7202 |
| $Hf_2$@4Nb | 47.259 | 2.6465 | 6.6635 | 11.7202 |
| $Hf_2$@4Nc | 47.259 | 2.6465 | 6.6635 | 11.7202 |
| $Hf_2$@4Nd | 47.259 | 2.6465 | 6.6635 | 11.7202 |
| $Hf_2$@3Na | 47.354 | 2.6518 | 6.6770 | 11.7439 |
| $Hf_2$@3Nb | 47.354 | 2.6518 | 6.6770 | 11.7439 |
| $Hf_2$@3Nc | 47.354 | 2.6518 | 6.6770 | 11.7439 |
| $Hf_2$@3Nd | 47.354 | 2.6518 | 6.6770 | 11.7439 |
| $Hf_2$@2Na | 47.450 | 2.6572 | 6.6905 | 11.7677 |
| $Hf_2$@2Nb | 47.450 | 2.6572 | 6.6905 | 11.7677 |
| $Hf_2$@2Nc | 47.450 | 2.6572 | 6.6905 | 11.7677 |
| $Hf_2$@2Nd | 47.450 | 2.6572 | 6.6905 | 11.7677 |
| $Hf_2$@2Ne | 47.450 | 2.6572 | 6.6905 | 11.7677 |
| $Hf_2$@2Nf | 47.450 | 2.6572 | 6.6905 | 11.7677 |
| $Hf_2$@1Na | 47.547 | 2.6626 | 6.7041 | 11.7916 |
| $Hf_2$@1Nb | 47.547 | 2.6626 | 6.7041 | 11.7916 |
| $Hf_2$@0N | 47.643 | 2.6680 | 6.7177 | 11.8155 |
| $Ag_2$@6N | 49.613 | 2.7783 | 6.9954 | 12.3040 |
| $Ag_2$@5Na | 49.729 | 2.7848 | 7.0118 | 12.3329 |
| $Ag_2$@5Nb | 49.729 | 2.7848 | 7.0118 | 12.3329 |
| $Ag_2$@4Na | 49.846 | 2.7914 | 7.0283 | 12.3619 |
| $Ag_2$@4Nb | 49.846 | 2.7914 | 7.0283 | 12.3619 |
| $Ag_2$@4Nc | 49.846 | 2.7914 | 7.0283 | 12.3619 |
| $Ag_2$@4Nd | 49.846 | 2.7914 | 7.0283 | 12.3619 |
| $Ag_2$@3Na | 49.964 | 2.7980 | 7.0449 | 12.3910 |
| $Ag_2$@3Nb | 49.964 | 2.7980 | 7.0449 | 12.3910 |
| $Ag_2$@3Nc | 49.964 | 2.7980 | 7.0449 | 12.3910 |
| $Ag_2$@3Nd | 49.964 | 2.7980 | 7.0449 | 12.3910 |
| $Ag_2$@2Na | 50.082 | 2.8046 | 7.0616 | 12.4203 |
| $Ag_2$@2Nb | 50.082 | 2.8046 | 7.0616 | 12.4203 |
| $Ag_2$@2Nc | 50.082 | 2.8046 | 7.0616 | 12.4203 |
| $Ag_2$@2Nd | 50.082 | 2.8046 | 7.0616 | 12.4203 |
| $Ag_2$@2Ne | 50.082 | 2.8046 | 7.0616 | 12.4203 |
| $Ag_2$@2Nf | 50.082 | 2.8046 | 7.0616 | 12.4203 |
| $Ag_2$@1Na | 50.201 | 2.8112 | 7.0783 | 12.4498 |
| $Ag_2$@1Nb | 50.201 | 2.8112 | 7.0783 | 12.4498 |
| $Ag_2$@0N | 50.320 | 2.8179 | 7.0951 | 12.4793 |
| $Ta_2$@6N | 94.240 | 5.2774 | 13.2878 | 23.3714 |
| $Ta_2$@5Na | 94.428 | 5.2880 | 13.3144 | 23.4182 |
| $Ta_2$@5Nb | 94.428 | 5.2880 | 13.3144 | 23.4182 |
| $Ta_2$@4Na | 94.618 | 5.2986 | 13.3411 | 23.4652 |
| $Ta_2$@4Nb | 94.618 | 5.2986 | 13.3411 | 23.4652 |
| $Ta_2$@4Nc | 94.618 | 5.2986 | 13.3411 | 23.4652 |
| $Ta_2$@4Nd | 94.618 | 5.2986 | 13.3411 | 23.4652 |
| $Ta_2$@3Na | 94.808 | 5.3093 | 13.3680 | 23.5124 |
| $Ta_2$@3Nb | 94.808 | 5.3093 | 13.3680 | 23.5124 |
| $Ta_2$@3Nc | 94.808 | 5.3093 | 13.3680 | 23.5124 |
| $Ta_2$@3Nd | 94.808 | 5.3093 | 13.3680 | 23.5124 |
| $Ta_2$@2Na | 94.999 | 5.3200 | 13.3949 | 23.5598 |
| $Ta_2$@2Nb | 94.999 | 5.3200 | 13.3949 | 23.5598 |
| $Ta_2$@2Nc | 94.999 | 5.3200 | 13.3949 | 23.5598 |
| $Ta_2$@2Nd | 94.999 | 5.3200 | 13.3949 | 23.5598 |
| $Ta_2$@2Ne | 94.999 | 5.3200 | 13.3949 | 23.5598 |
| $Ta_2$@2Nf | 94.999 | 5.3200 | 13.3949 | 23.5598 |
| $Ta_2$@1Na | 95.191 | 5.3307 | 13.4219 | 23.6074 |
| $Ta_2$@1Nb | 95.191 | 5.3307 | 13.4219 | 23.6074 |
| $Ta_2$@0N | 95.384 | 5.3415 | 13.4491 | 23.6552 |
| $Re_2$@6N | 166.098 | 9.3015 | 23.4198 | 41.1922 |
| $Re_2$@5Na | 166.427 | 9.3199 | 23.4662 | 41.2739 |
| $Re_2$@5Nb | 166.427 | 9.3199 | 23.4662 | 41.2739 |
| $Re_2$@4Na | 166.758 | 9.3384 | 23.5128 | 41.3559 |
| $Re_2$@4Nb | 166.758 | 9.3384 | 23.5128 | 41.3559 |
| $Re_2$@4Nc | 166.758 | 9.3384 | 23.5128 | 41.3559 |
| $Re_2$@4Nd | 166.758 | 9.3384 | 23.5128 | 41.3559 |
| $Re_2$@3Na | 167.089 | 9.3570 | 23.5596 | 41.4382 |
| $Re_2$@3Nb | 167.089 | 9.3570 | 23.5596 | 41.4382 |
| $Re_2$@3Nc | 167.089 | 9.3570 | 23.5596 | 41.4382 |
| $Re_2$@3Nd | 167.089 | 9.3570 | 23.5596 | 41.4382 |
| $Re_2$@2Na | 167.423 | 9.3757 | 23.6066 | 41.5208 |
| $Re_2$@2Nb | 167.423 | 9.3757 | 23.6066 | 41.5208 |
| $Re_2$@2Nc | 167.423 | 9.3757 | 23.6066 | 41.5208 |
| $Re_2$@2Nd | 167.423 | 9.3757 | 23.6066 | 41.5208 |
| $Re_2$@2Ne | 167.423 | 9.3757 | 23.6066 | 41.5208 |
| $Re_2$@2Nf | 167.423 | 9.3757 | 23.6066 | 41.5208 |
| $Re_2$@1Na | 167.757 | 9.3944 | 23.6537 | 41.6038 |
| $Re_2$@1Nb | 167.757 | 9.3944 | 23.6537 | 41.6038 |
| $Re_2$@0N | 168.093 | 9.4132 | 23.7011 | 41.6870 |
| $Ru_2$@6N | 508.551 | 28.4788 | 71.7057 | 126.1 |
| $Ru_2$@5Na | 509.764 | 28.5468 | 71.8767 | 126.4 |
| $Ru_2$@5Nb | 509.764 | 28.5468 | 71.8767 | 126.4 |
| $Ru_2$@4Na | 510.983 | 28.6150 | 72.0486 | 126.7 |
| $Ru_2$@4Nb | 510.983 | 28.6150 | 72.0486 | 126.7 |
| $Ru_2$@4Nc | 510.983 | 28.6150 | 72.0486 | 126.7 |
| $Ru_2$@4Nd | 510.983 | 28.6150 | 72.0486 | 126.7 |
| $Ru_2$@3Na | 512.208 | 28.6836 | 72.2213 | 127.0 |
| $Ru_2$@3Nb | 512.208 | 28.6836 | 72.2213 | 127.0 |
| $Ru_2$@3Nc | 512.208 | 28.6836 | 72.2213 | 127.0 |
| $Ru_2$@3Nd | 512.208 | 28.6836 | 72.2213 | 127.0 |
| $Ru_2$@2Na | 513.438 | 28.7526 | 72.3948 | 127.3 |
| $Ru_2$@2Nb | 513.438 | 28.7526 | 72.3948 | 127.3 |
| $Ru_2$@2Nc | 513.438 | 28.7526 | 72.3948 | 127.3 |
| $Ru_2$@2Nd | 513.438 | 28.7526 | 72.3948 | 127.3 |
| $Ru_2$@2Ne | 513.438 | 28.7526 | 72.3948 | 127.3 |
| $Ru_2$@2Nf | 513.438 | 28.7526 | 72.3948 | 127.3 |
| $Ru_2$@1Na | 514.675 | 28.8218 | 72.5692 | 127.6 |
| $Ru_2$@1Nb | 514.675 | 28.8218 | 72.5692 | 127.6 |
| $Ru_2$@0N | 515.918 | 28.8914 | 72.7444 | 127.9 |
| $Pd_2$@6N | 972.252 | 54.4461 | 137.1 | 241.1 |
| $Pd_2$@5Na | 974.542 | 54.5744 | 137.4 | 241.7 |
| $Pd_2$@5Nb | 974.542 | 54.5744 | 137.4 | 241.7 |
| $Pd_2$@4Na | 976.843 | 54.7032 | 137.7 | 242.3 |
| $Pd_2$@4Nb | 976.843 | 54.7032 | 137.7 | 242.3 |
| $Pd_2$@4Nc | 976.843 | 54.7032 | 137.7 | 242.3 |
| $Pd_2$@4Nd | 976.843 | 54.7032 | 137.7 | 242.3 |
| $Pd_2$@3Na | 979.155 | 54.8327 | 138.1 | 242.8 |
| $Pd_2$@3Nb | 979.155 | 54.8327 | 138.1 | 242.8 |
| $Pd_2$@3Nc | 979.155 | 54.8327 | 138.1 | 242.8 |
| $Pd_2$@3Nd | 979.155 | 54.8327 | 138.1 | 242.8 |
| $Pd_2$@2Na | 981.478 | 54.9627 | 138.4 | 243.4 |
| $Pd_2$@2Nb | 981.478 | 54.9627 | 138.4 | 243.4 |
| $Pd_2$@2Nc | 981.478 | 54.9627 | 138.4 | 243.4 |
| $Pd_2$@2Nd | 981.478 | 54.9627 | 138.4 | 243.4 |
| $Pd_2$@2Ne | 981.478 | 54.9627 | 138.4 | 243.4 |
| $Pd_2$@2Nf | 981.478 | 54.9627 | 138.4 | 243.4 |
| $Pd_2$@1Na | 983.811 | 55.0934 | 138.7 | 244.0 |
| $Pd_2$@1Nb | 983.811 | 55.0934 | 138.7 | 244.0 |
| $Pd_2$@0N | 986.156 | 55.2248 | 139.0 | 244.6 |
| $Os_2$@6N | 1,705.884 | 95.5295 | 240.5 | 423.1 |
| $Os_2$@5Na | 1,709.239 | 95.7174 | 241.0 | 423.9 |
| $Os_2$@5Nb | 1,709.239 | 95.7174 | 241.0 | 423.9 |
| $Os_2$@4Na | 1,712.606 | 95.9060 | 241.5 | 424.7 |
| $Os_2$@4Nb | 1,712.606 | 95.9060 | 241.5 | 424.7 |
| $Os_2$@4Nc | 1,712.606 | 95.9060 | 241.5 | 424.7 |
| $Os_2$@4Nd | 1,712.606 | 95.9060 | 241.5 | 424.7 |
| $Os_2$@3Na | 1,715.987 | 96.0953 | 242.0 | 425.6 |
| $Os_2$@3Nb | 1,715.987 | 96.0953 | 242.0 | 425.6 |
| $Os_2$@3Nc | 1,715.987 | 96.0953 | 242.0 | 425.6 |
| $Os_2$@3Nd | 1,715.987 | 96.0953 | 242.0 | 425.6 |
| $Os_2$@2Na | 1,719.382 | 96.2854 | 242.4 | 426.4 |
| $Os_2$@2Nb | 1,719.382 | 96.2854 | 242.4 | 426.4 |
| $Os_2$@2Nc | 1,719.382 | 96.2854 | 242.4 | 426.4 |
| $Os_2$@2Nd | 1,719.382 | 96.2854 | 242.4 | 426.4 |
| $Os_2$@2Ne | 1,719.382 | 96.2854 | 242.4 | 426.4 |
| $Os_2$@2Nf | 1,719.382 | 96.2854 | 242.4 | 426.4 |
| $Os_2$@1Na | 1,722.790 | 96.4762 | 242.9 | 427.3 |
| $Os_2$@1Nb | 1,722.790 | 96.4762 | 242.9 | 427.3 |
| $Os_2$@0N | 1,726.211 | 96.6678 | 243.4 | 428.1 |
| $Ir_2$@6N | 3,336.087 | 186.8 | 470.4 | 827.3 |
| $Ir_2$@5Na | 3,342.622 | 187.2 | 471.3 | 829.0 |
| $Ir_2$@5Nb | 3,342.622 | 187.2 | 471.3 | 829.0 |
| $Ir_2$@4Na | 3,349.183 | 187.6 | 472.2 | 830.6 |
| $Ir_2$@4Nb | 3,349.183 | 187.6 | 472.2 | 830.6 |
| $Ir_2$@4Nc | 3,349.183 | 187.6 | 472.2 | 830.6 |
| $Ir_2$@4Nd | 3,349.183 | 187.6 | 472.2 | 830.6 |
| $Ir_2$@3Na | 3,355.769 | 187.9 | 473.2 | 832.2 |
| $Ir_2$@3Nb | 3,355.769 | 187.9 | 473.2 | 832.2 |
| $Ir_2$@3Nc | 3,355.769 | 187.9 | 473.2 | 832.2 |
| $Ir_2$@3Nd | 3,355.769 | 187.9 | 473.2 | 832.2 |
| $Ir_2$@2Na | 3,362.381 | 188.3 | 474.1 | 833.9 |
| $Ir_2$@2Nb | 3,362.381 | 188.3 | 474.1 | 833.9 |
| $Ir_2$@2Nc | 3,362.381 | 188.3 | 474.1 | 833.9 |
| $Ir_2$@2Nd | 3,362.381 | 188.3 | 474.1 | 833.9 |
| $Ir_2$@2Ne | 3,362.381 | 188.3 | 474.1 | 833.9 |
| $Ir_2$@2Nf | 3,362.381 | 188.3 | 474.1 | 833.9 |
| $Ir_2$@1Na | 3,369.019 | 188.7 | 475.0 | 835.5 |
| $Ir_2$@1Nb | 3,369.019 | 188.7 | 475.0 | 835.5 |
| $Ir_2$@0N | 3,375.684 | 189.0 | 476.0 | 837.2 |
| $Pt_2$@6N | 4,750.194 | 266.0 | 669.8 | 1,178.0 |
| $Pt_2$@5Na | 4,759.446 | 266.5 | 671.1 | 1,180.3 |
| $Pt_2$@5Nb | 4,759.446 | 266.5 | 671.1 | 1,180.3 |
| $Pt_2$@4Na | 4,768.735 | 267.0 | 672.4 | 1,182.6 |
| $Pt_2$@4Nb | 4,768.735 | 267.0 | 672.4 | 1,182.6 |
| $Pt_2$@4Nc | 4,768.735 | 267.0 | 672.4 | 1,182.6 |
| $Pt_2$@4Nd | 4,768.735 | 267.0 | 672.4 | 1,182.6 |
| $Pt_2$@3Na | 4,778.060 | 267.6 | 673.7 | 1,185.0 |
| $Pt_2$@3Nb | 4,778.060 | 267.6 | 673.7 | 1,185.0 |
| $Pt_2$@3Nc | 4,778.060 | 267.6 | 673.7 | 1,185.0 |
| $Pt_2$@3Nd | 4,778.060 | 267.6 | 673.7 | 1,185.0 |
| $Au_2$@6N | 4,778.516 | 267.6 | 673.8 | 1,185.1 |
| $Pt_2$@2Na | 4,787.422 | 268.1 | 675.0 | 1,187.3 |
| $Pt_2$@2Nb | 4,787.422 | 268.1 | 675.0 | 1,187.3 |
| $Pt_2$@2Nc | 4,787.422 | 268.1 | 675.0 | 1,187.3 |
| $Pt_2$@2Nd | 4,787.422 | 268.1 | 675.0 | 1,187.3 |
| $Pt_2$@2Ne | 4,787.422 | 268.1 | 675.0 | 1,187.3 |
| $Pt_2$@2Nf | 4,787.422 | 268.1 | 675.0 | 1,187.3 |
| $Au_2$@5Na | 4,787.790 | 268.1 | 675.1 | 1,187.4 |
| $Au_2$@5Nb | 4,787.790 | 268.1 | 675.1 | 1,187.4 |
| $Pt_2$@1Na | 4,796.820 | 268.6 | 676.4 | 1,189.6 |
| $Pt_2$@1Nb | 4,796.820 | 268.6 | 676.4 | 1,189.6 |
| $Au_2$@4Na | 4,797.100 | 268.6 | 676.4 | 1,189.7 |
| $Au_2$@4Nb | 4,797.100 | 268.6 | 676.4 | 1,189.7 |
| $Au_2$@4Nc | 4,797.100 | 268.6 | 676.4 | 1,189.7 |
| $Au_2$@4Nd | 4,797.100 | 268.6 | 676.4 | 1,189.7 |
| $Pt_2$@0N | 4,806.256 | 269.2 | 677.7 | 1,192.0 |
| $Au_2$@3Na | 4,806.446 | 269.2 | 677.7 | 1,192.0 |
| $Au_2$@3Nb | 4,806.446 | 269.2 | 677.7 | 1,192.0 |
| $Au_2$@3Nc | 4,806.446 | 269.2 | 677.7 | 1,192.0 |
| $Au_2$@3Nd | 4,806.446 | 269.2 | 677.7 | 1,192.0 |
| $Au_2$@2Na | 4,815.828 | 269.7 | 679.0 | 1,194.3 |
| $Au_2$@2Nb | 4,815.828 | 269.7 | 679.0 | 1,194.3 |
| $Au_2$@2Nc | 4,815.828 | 269.7 | 679.0 | 1,194.3 |
| $Au_2$@2Nd | 4,815.828 | 269.7 | 679.0 | 1,194.3 |
| $Au_2$@2Ne | 4,815.828 | 269.7 | 679.0 | 1,194.3 |
| $Au_2$@2Nf | 4,815.828 | 269.7 | 679.0 | 1,194.3 |
| $Au_2$@1Na | 4,825.248 | 270.2 | 680.4 | 1,196.7 |
| $Au_2$@1Nb | 4,825.248 | 270.2 | 680.4 | 1,196.7 |
| $Au_2$@0N | 4,834.704 | 270.7 | 681.7 | 1,199.0 |
| $Rh_2$@6N | 8,575.907 | 480.3 | 1,209.2 | 2,126.8 |
| $Rh_2$@5Na | 8,596.276 | 481.4 | 1,212.1 | 2,131.9 |
| $Rh_2$@5Nb | 8,596.276 | 481.4 | 1,212.1 | 2,131.9 |
| $Rh_2$@4Na | 8,616.743 | 482.5 | 1,215.0 | 2,137.0 |
| $Rh_2$@4Nb | 8,616.743 | 482.5 | 1,215.0 | 2,137.0 |
| $Rh_2$@4Nc | 8,616.743 | 482.5 | 1,215.0 | 2,137.0 |
| $Rh_2$@4Nd | 8,616.743 | 482.5 | 1,215.0 | 2,137.0 |
| $Rh_2$@3Na | 8,637.306 | 483.7 | 1,217.9 | 2,142.1 |
| $Rh_2$@3Nb | 8,637.306 | 483.7 | 1,217.9 | 2,142.1 |
| $Rh_2$@3Nc | 8,637.306 | 483.7 | 1,217.9 | 2,142.1 |
| $Rh_2$@3Nd | 8,637.306 | 483.7 | 1,217.9 | 2,142.1 |
| $Rh_2$@2Na | 8,657.969 | 484.8 | 1,220.8 | 2,147.2 |
| $Rh_2$@2Nb | 8,657.969 | 484.8 | 1,220.8 | 2,147.2 |
| $Rh_2$@2Nc | 8,657.969 | 484.8 | 1,220.8 | 2,147.2 |
| $Rh_2$@2Nd | 8,657.969 | 484.8 | 1,220.8 | 2,147.2 |
| $Rh_2$@2Ne | 8,657.969 | 484.8 | 1,220.8 | 2,147.2 |
| $Rh_2$@2Nf | 8,657.969 | 484.8 | 1,220.8 | 2,147.2 |
| $Rh_2$@1Na | 8,678.730 | 486.0 | 1,223.7 | 2,152.3 |
| $Rh_2$@1Nb | 8,678.730 | 486.0 | 1,223.7 | 2,152.3 |
| $Rh_2$@0N | 8,699.591 | 487.2 | 1,226.6 | 2,157.5 |

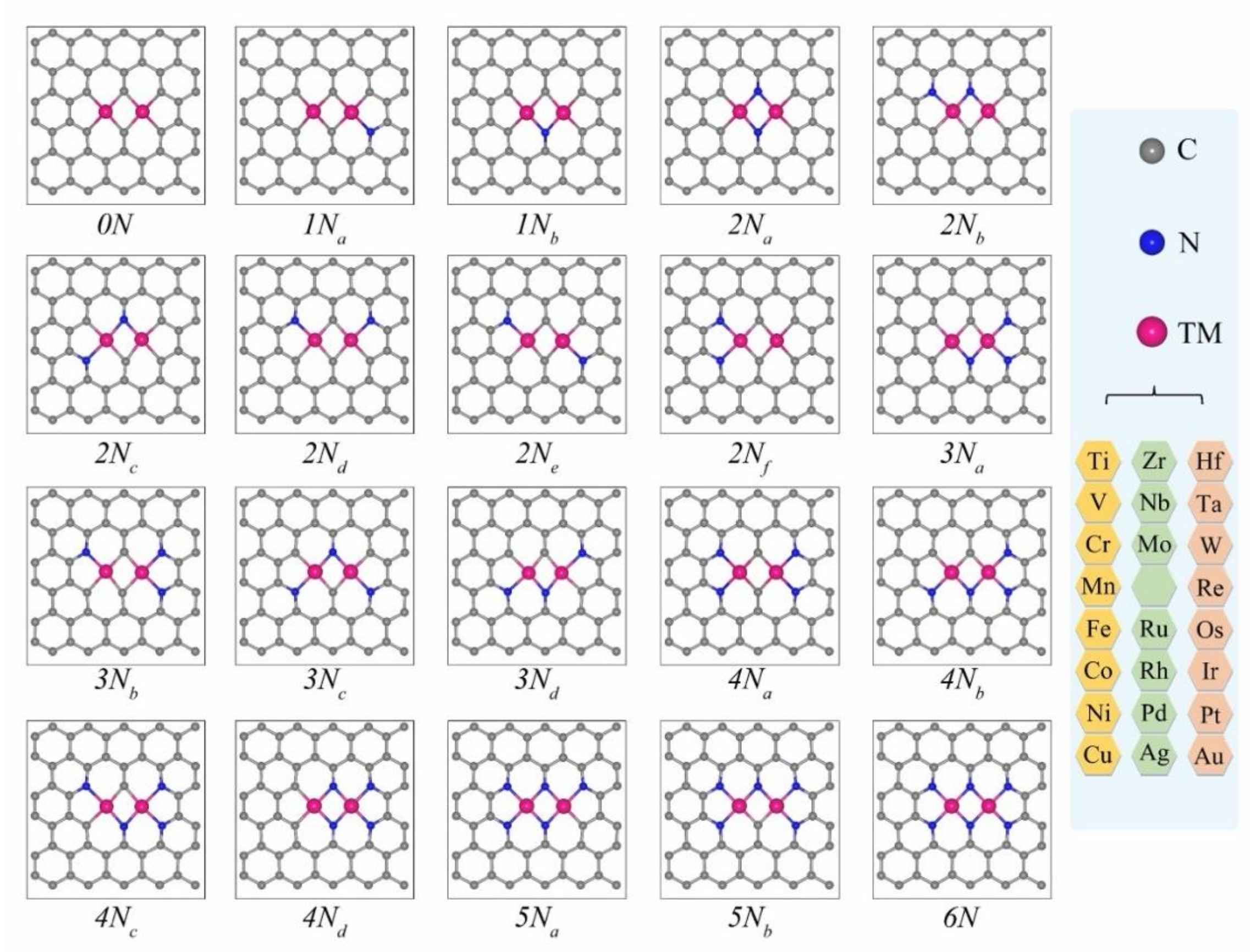

**Figure S1**. Schematic illustration of nitrogen coordinated, graphene supported dual atom catalysts ($TM_2$@$N_x$-Gr) investigated in this work. The 20 distinct coordination environments are considered, ranging from pristine double vacancy (0N) through 1N, 2N, 3N, 4N, and 5N motifs to fully coordinated 6N, with multiple structural isomers (denoted by subscripts *a–f*) capturing different local geometric arrangements of the coordinating nitrogen atoms around the $TM_2$ dimer. Gray, blue, and magenta spheres represent C, N, and transition metal (TM) atoms, respectively. The 23 TMs screened span the 3d (Ti, V, Cr, Mn, Fe, Co, Ni, Cu), 4d (Zr, Nb, Mo, Ru, Rh, Pd, Ag), and 5d (Hf, Ta, W, Re, Os, Ir, Pt, Au) series, yielding a total of 460 homonuclear $TM_2$@$N_x$-Gr configurations.

Recently, nitrogen-doped graphene (NC) has emerged as highly promising substrates for atomically dispersed catalysts, owing to their tunable electronic structure, excellent electrical conductivity, large surface area, and robust metal–support interactions.[1-5] Compared with conventional 2D materials, nitrogen-doped graphene can effectively anchor isolated metal atoms or metal dimers through strong coordination interactions, thereby suppressing metal aggregation and improving structural stability under electrochemical conditions. These unique characteristics make nitrogen-doped graphene an ideal platform for constructing high-performance electrocatalysts for HER and other energy-conversion reactions.[3,6] Based on these advantages, a series of nitrogen-doped graphene-supported diatomic metal catalysts ($TM_2@N_x$–Gr) were systematically constructed (**Figure S1**), to elucidate the role of coordination environment in regulating catalytic performance. In these models, two TM atoms are embedded within graphene vacancies and coordinated by nitrogen atoms with different coordination numbers and local geometric configurations. The coordination environments range from 0N to 6N systems, and include multiple structural isomers (e.g. $1N_a$, $1N_b$, and $2N_a$-$2N_f$……), enabling a systematic investigation of coordination-dependent structure, electronic and catalytic properties. The introduction of nitrogen atoms not only stabilizes the dual-metal centers through strong metal-nitrogen interactions, but also significantly modifies the local electronic structure of the active sites. Specifically, nitrogen dopants act as donating centers depending on their local environment, effectively tuning the d-band center of the metal sites and thus altering the adsorption strength of key reaction intermediates. Moreover, variations in nitrogen coordination number and local atomic arrangement lead to distinct geometric symmetry and electronic coupling between the two metal centers. These differences can strongly influence charge-transfer behavior, orbital hybridization, and metal–metal interactions, ultimately affecting catalytic activity and reaction pathways. To ensure broad applicability and establish general coordination–activity relationships, a series of homonuclear diatomic electrocatalysts based on 3d, 4d, and 5d TMs were considered (**Figure S1**).

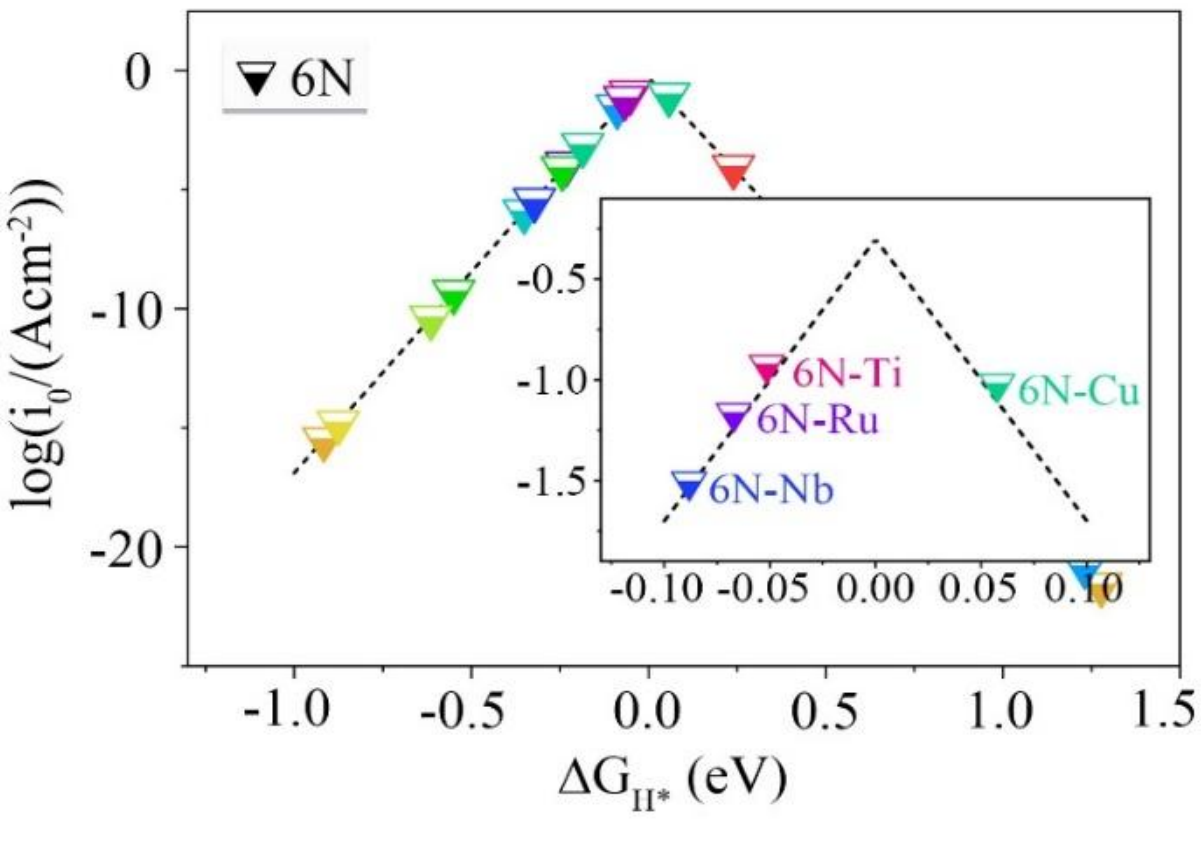

**Figure S2.** Volcano plots of the exchange current density ($i_0$) as a function of $\Delta G_{H*}$ for $TM_2@6N$-Gr.

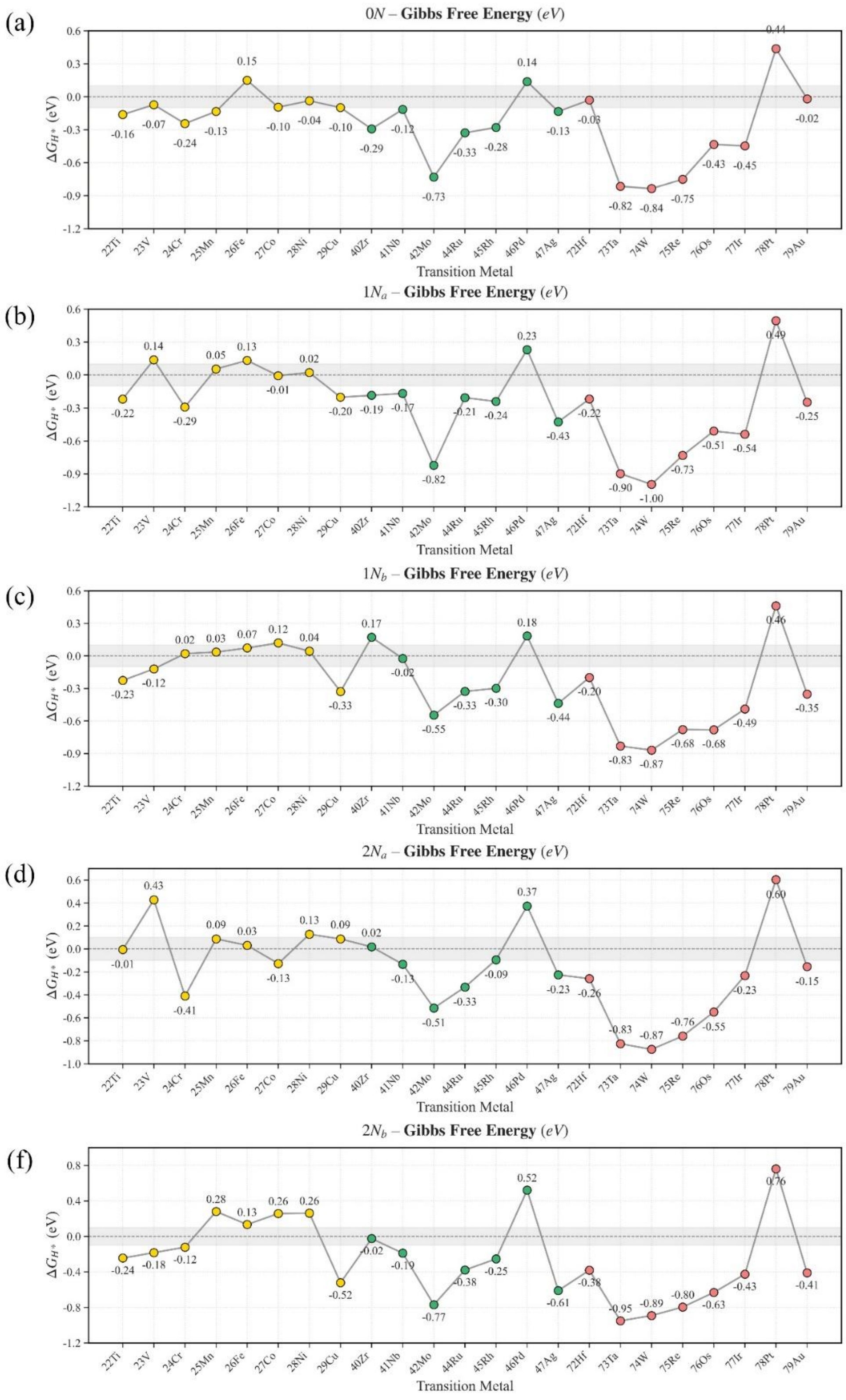

**Figure S3.** $\Delta G_{H*}$ for $TM_2@N_x$–Gr systems with varying nitrogen coordination. (a–f) correspond to different coordination environments: 0N, $1N_a$, $1N_b$, $2N_a$, and $2N_b$, respectively.

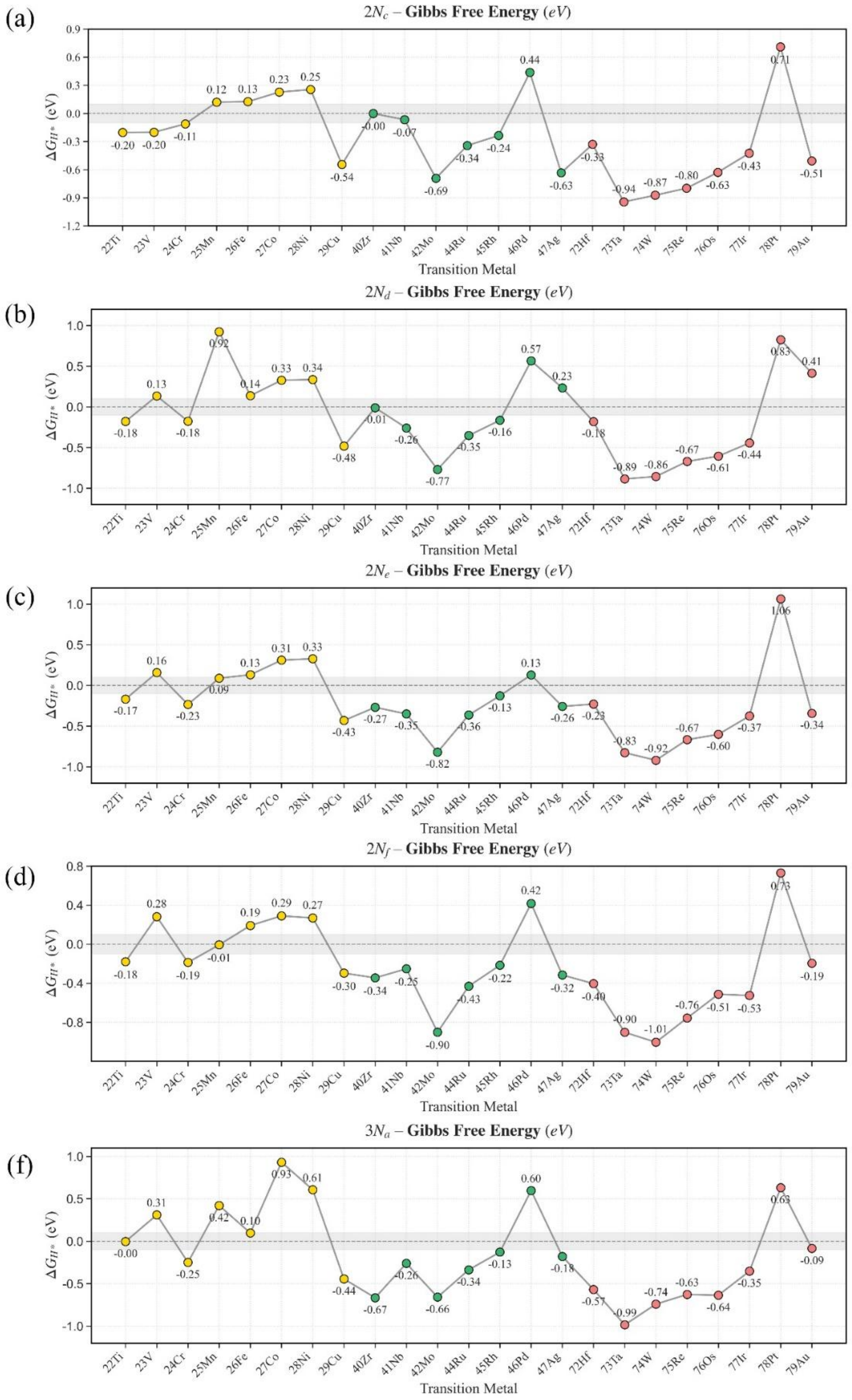

**Figure S4.** $\Delta G_{H^*}$ for $TM_2@N_x$–Gr systems with varying nitrogen coordination. (a–f) correspond to different coordination environments: $2N_c$, $2N_d$, $2N_e$, $2N_f$, and $3N_a$, respectively.

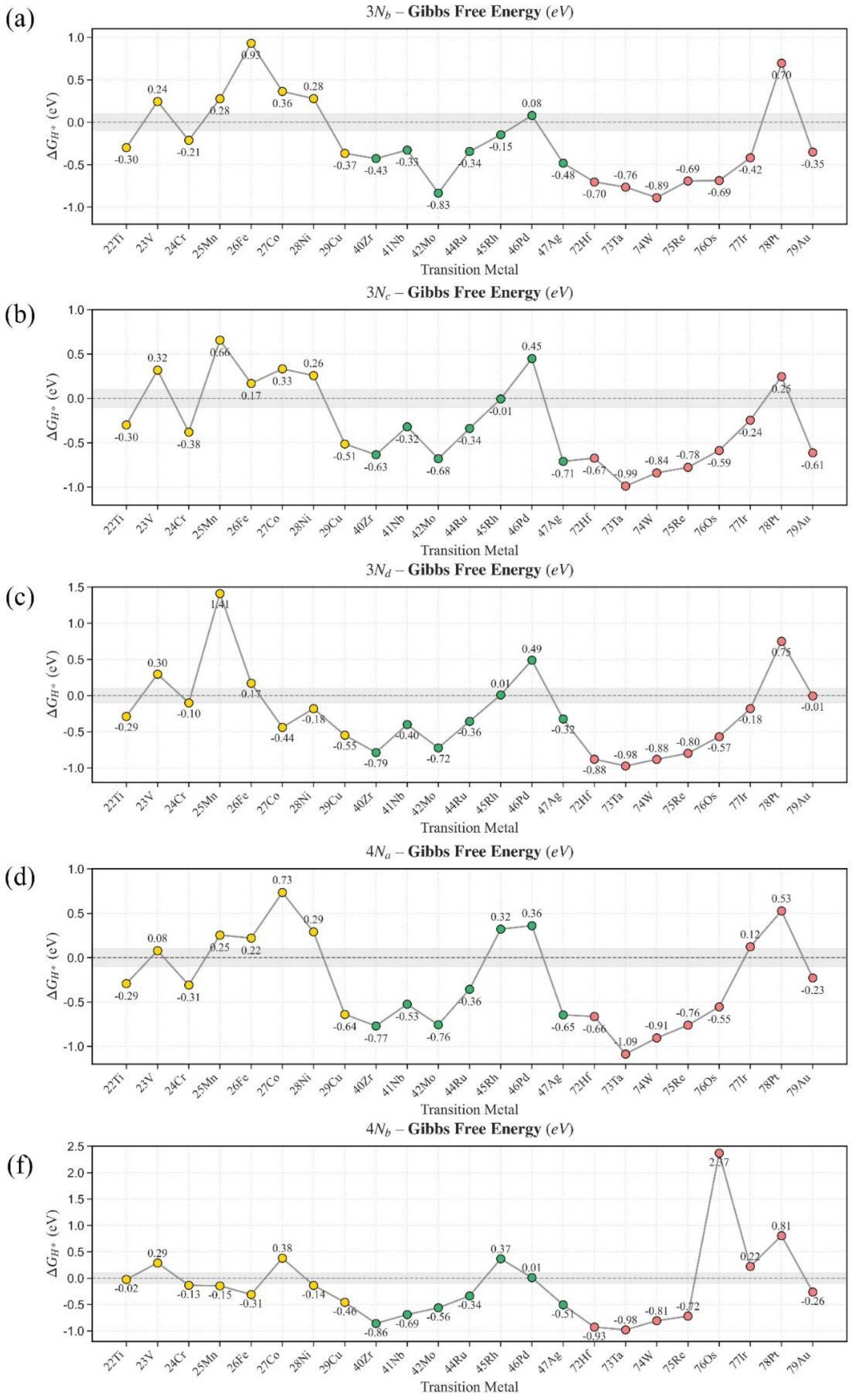

**Figure S5.** $\Delta G_{H^*}$ for $TM_2@N_x$–Gr systems with varying nitrogen coordination. (a–f) correspond to different coordination environments: $3N_b$, $3N_c$, $3N_d$, $4N_a$, and $4N_b$, respectively.

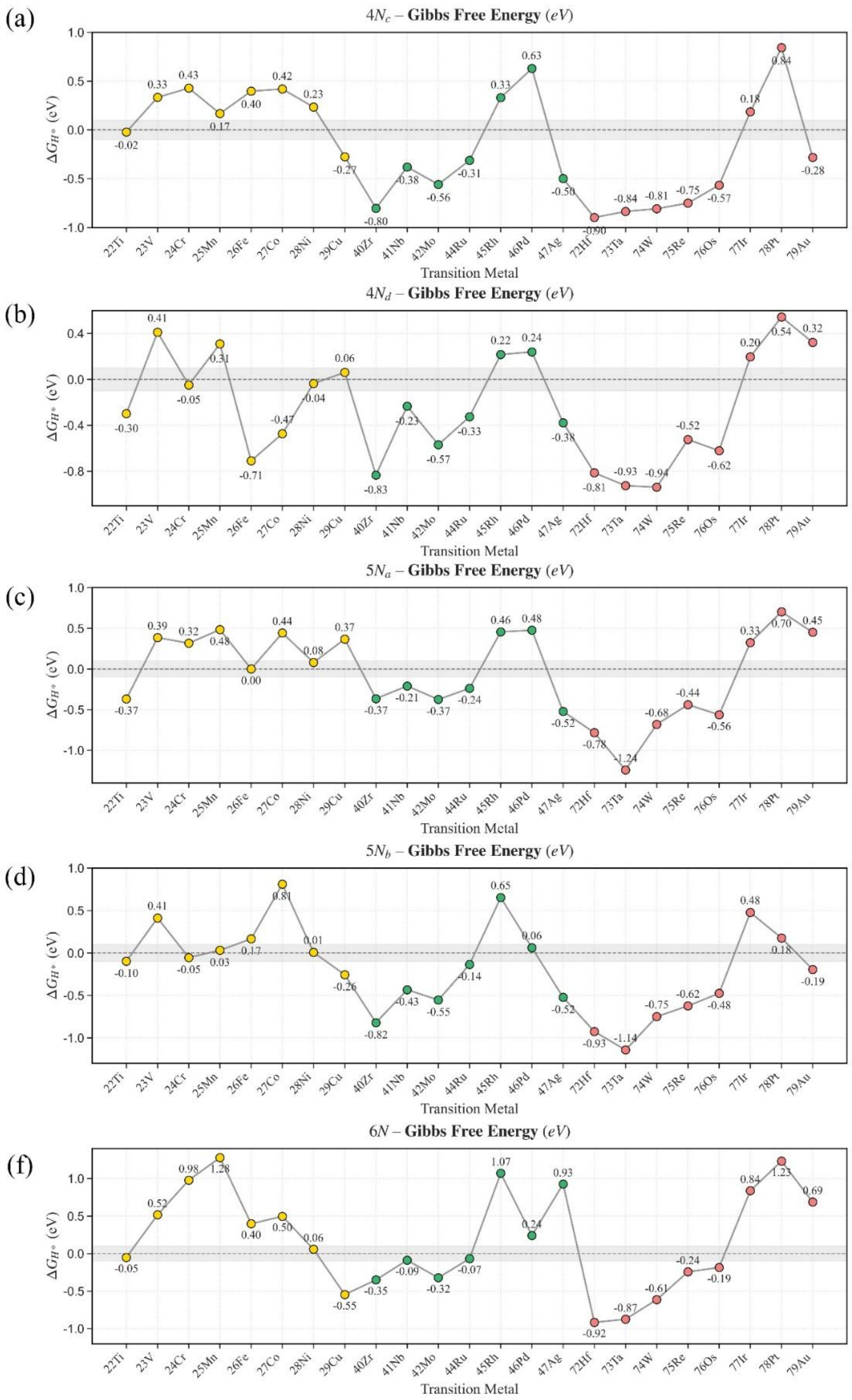

**Figure S6.** $\Delta G_{H*}$ for $TM_2@N_x$–Gr systems with varying nitrogen coordination. (a–f) correspond to different coordination environments: $4N_c$, $4N_d$, $5N_a$, $5N_b$, and 6N, respectively.

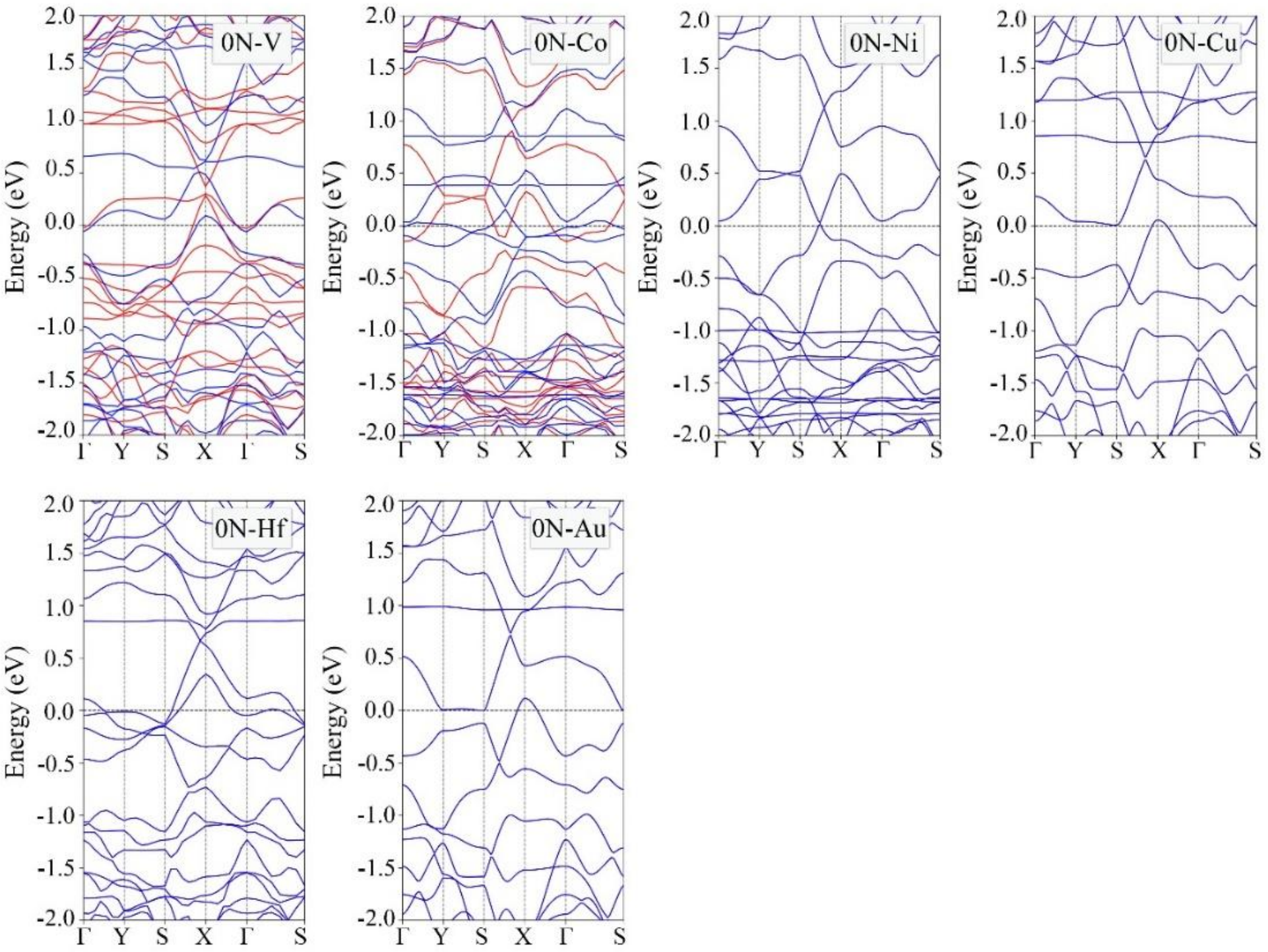

**Figure S7.** Band structures of $TM_2$@0N-Gr (TM = V, Co, Ni, Cu, Hf and Au). The red and blue lines represent spin-up and spin-down bands, respectively, and the Fermi level is set to 0 eV.

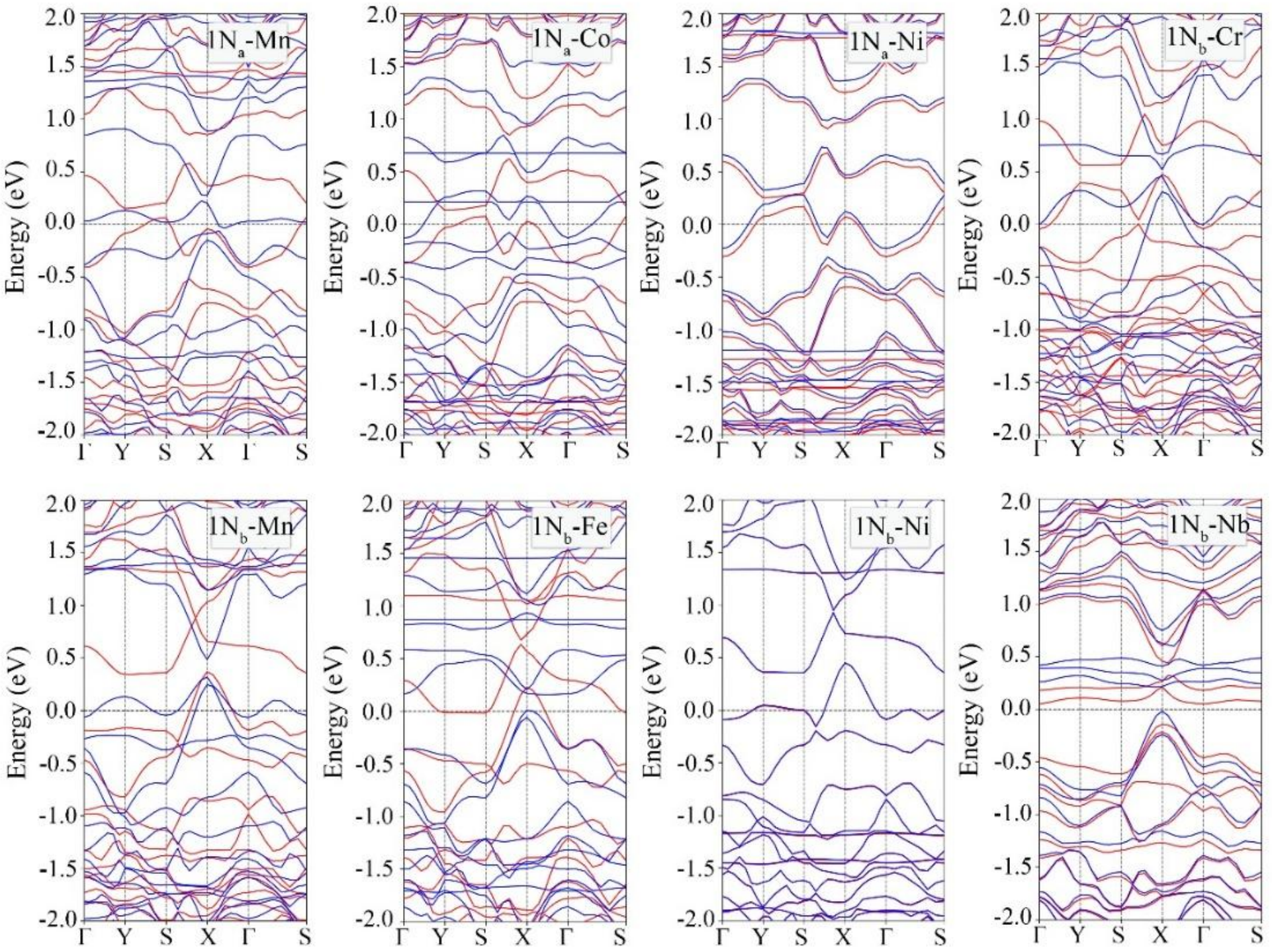

**Figure S8.** Band structures of $TM_2@1N_x$-Gr (x=a,b; TM = Mn, Co, Ni, Cr, Fe, and Nb). The red and blue lines represent spin-up and spin-down bands, respectively, and the Fermi level is set to 0 eV.

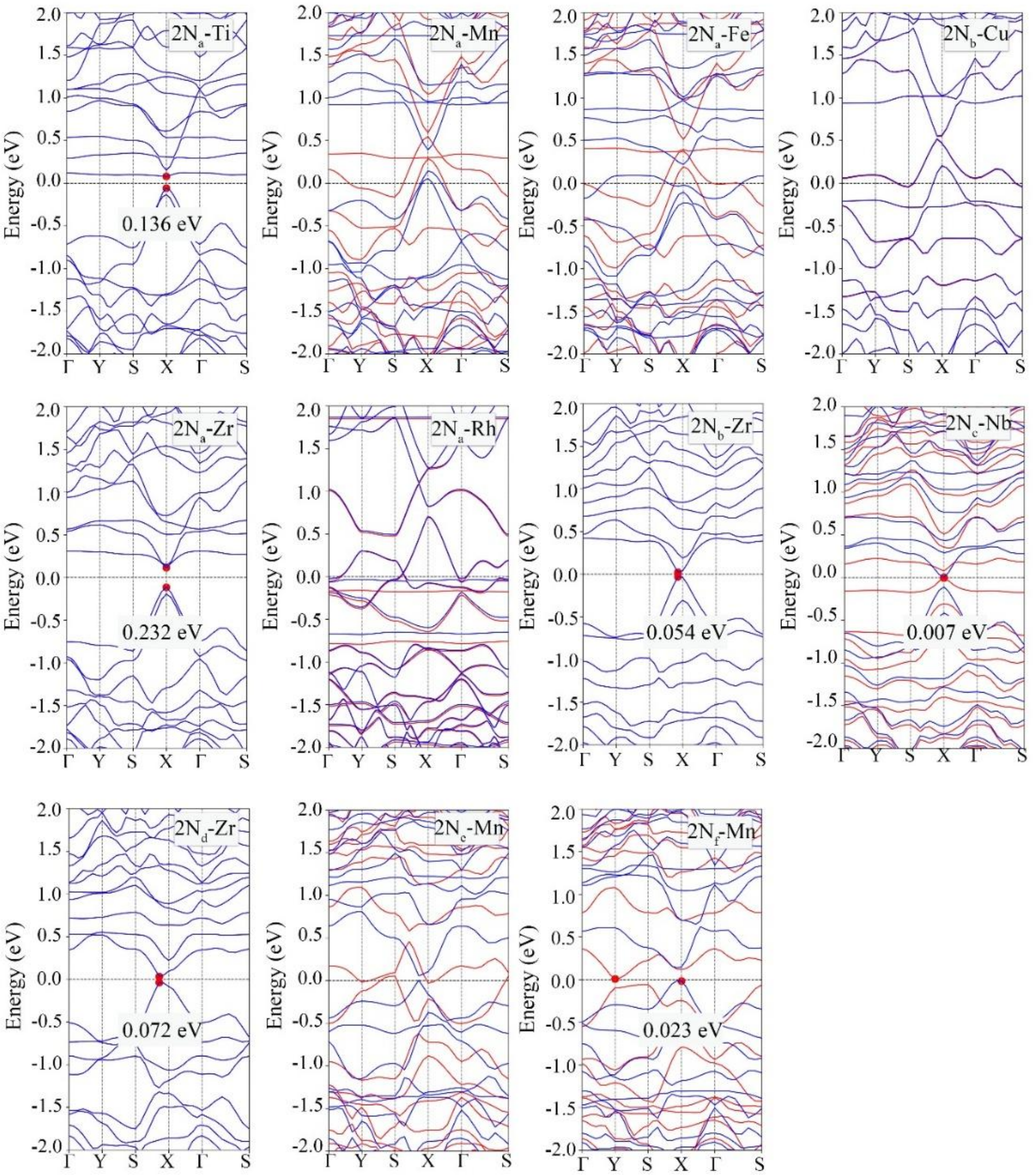

**Figure S9.** Band structures of $TM_2@2N_x$-Gr (x=a,b,c,d,e,f; TM = Ti, Mn, Fe, Cu, Zr, Rh, Zr, and Nb). The red and blue lines represent spin-up and spin-down bands, respectively, and the Fermi level is set to 0 eV.

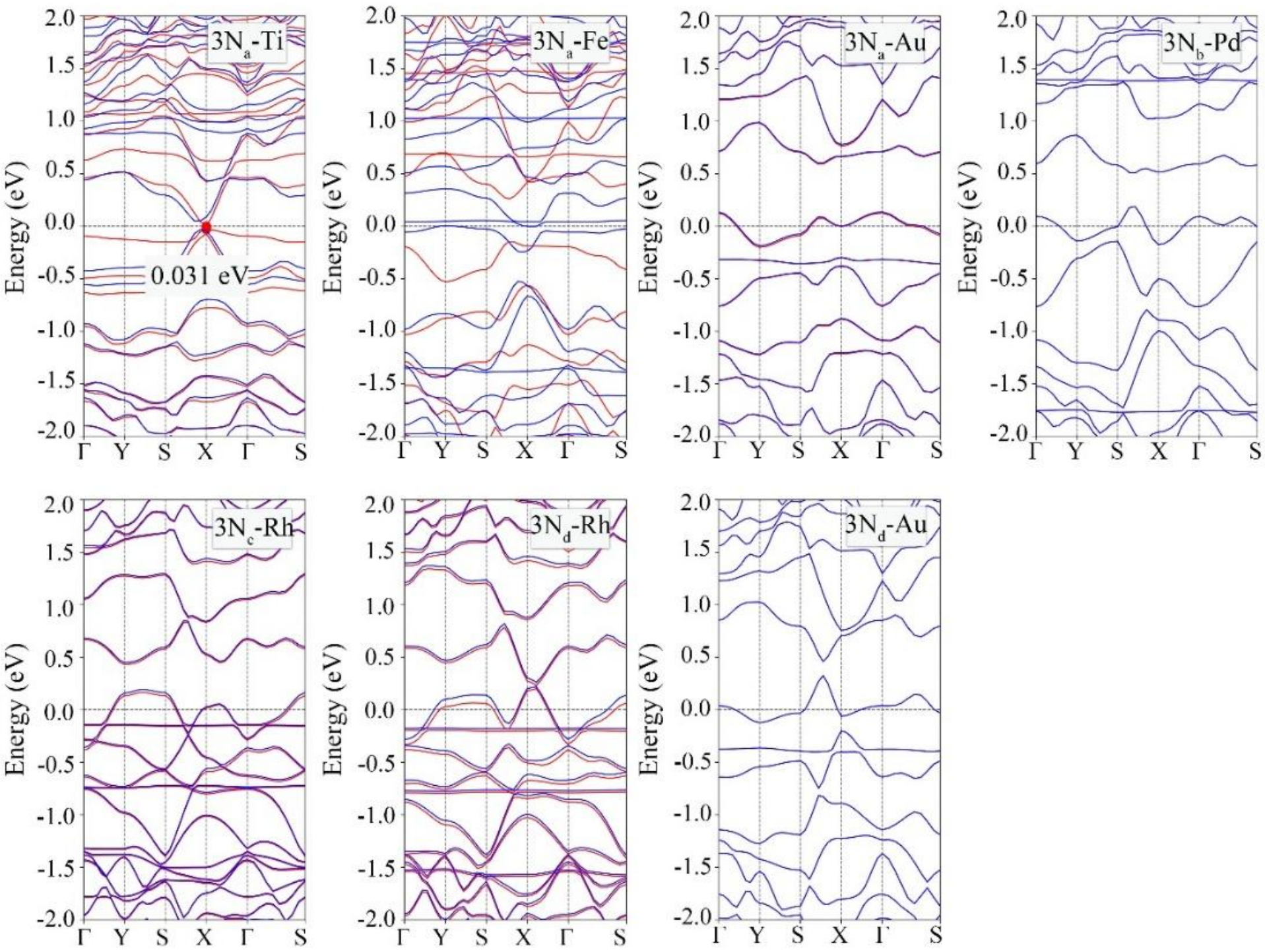

**Figure S10.** Band structures of $TM_2$@$3N_x$-Gr (x=a,b,c,d; TM = Ti, Fe, Au, Pd, Rh, and Au). The red and blue lines represent spin-up and spin-down bands, respectively, and the Fermi level is set to 0 eV.

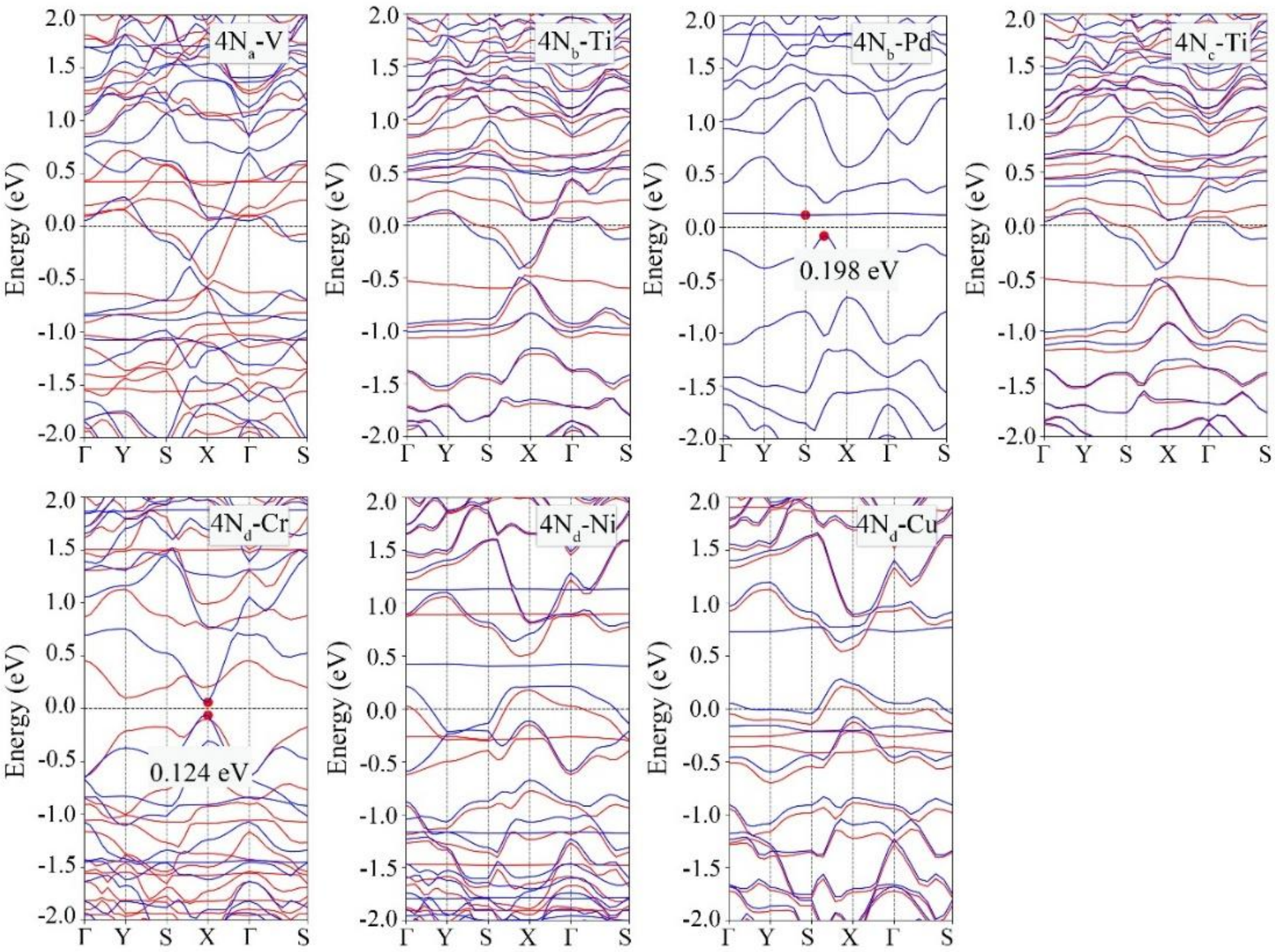

**Figure S11.** Band structures of $TM_2$@$4N_x$-Gr (x=a,b,c,d; TM = V, Ti, Pd, Cr, Ni, and Cu) The red and blue lines represent spin-up and spin-down bands, respectively, and the Fermi level is set to 0 eV.

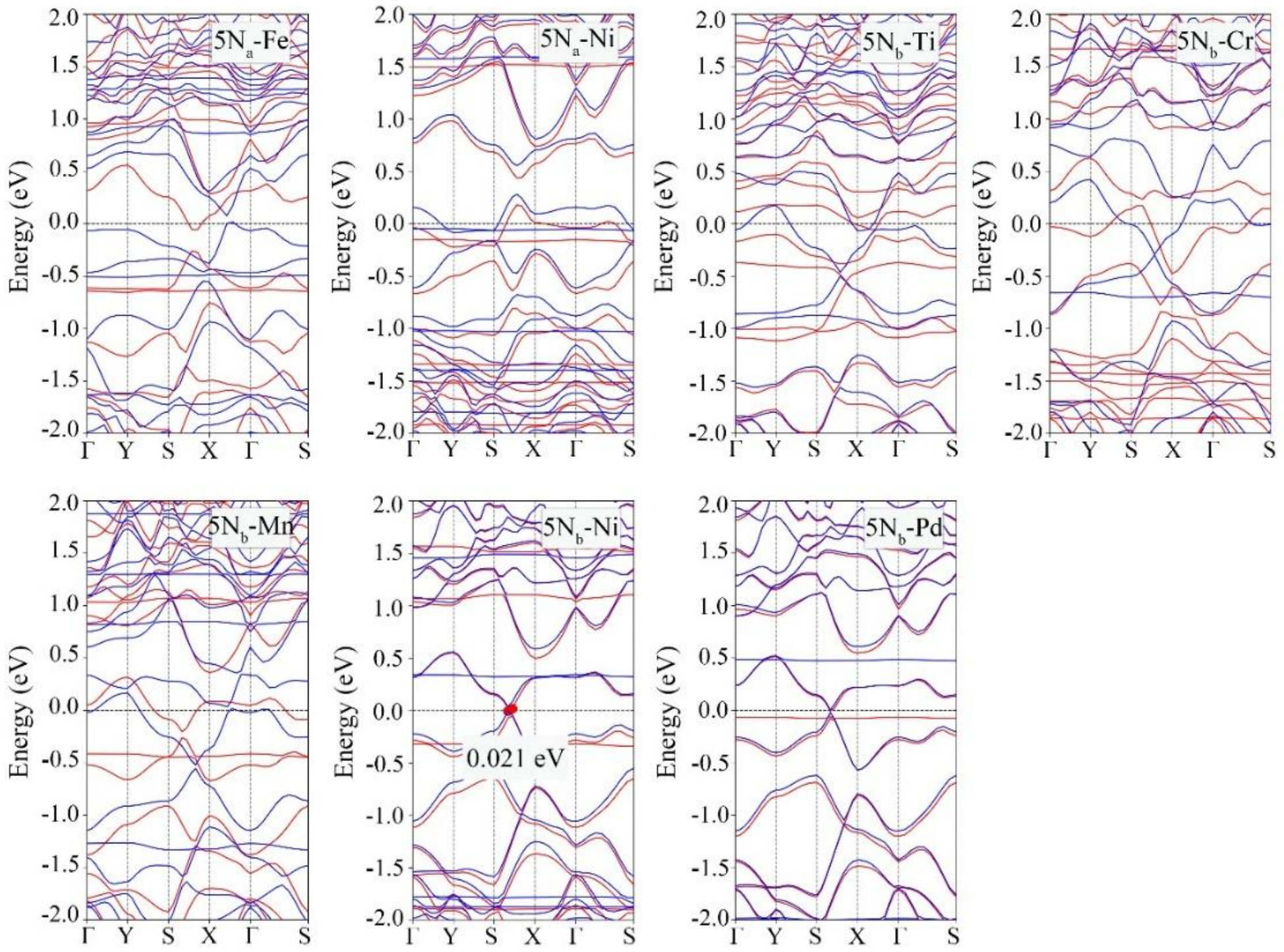

**Figure S12.** Band structures of $TM_2$@$5N_x$-Gr (x=a,b; TM = Fe, Ni, Ti, Cr, Mn, Ni and Pd). The red and blue lines represent spin-up and spin-down bands, respectively, and the Fermi level is set to 0 eV.

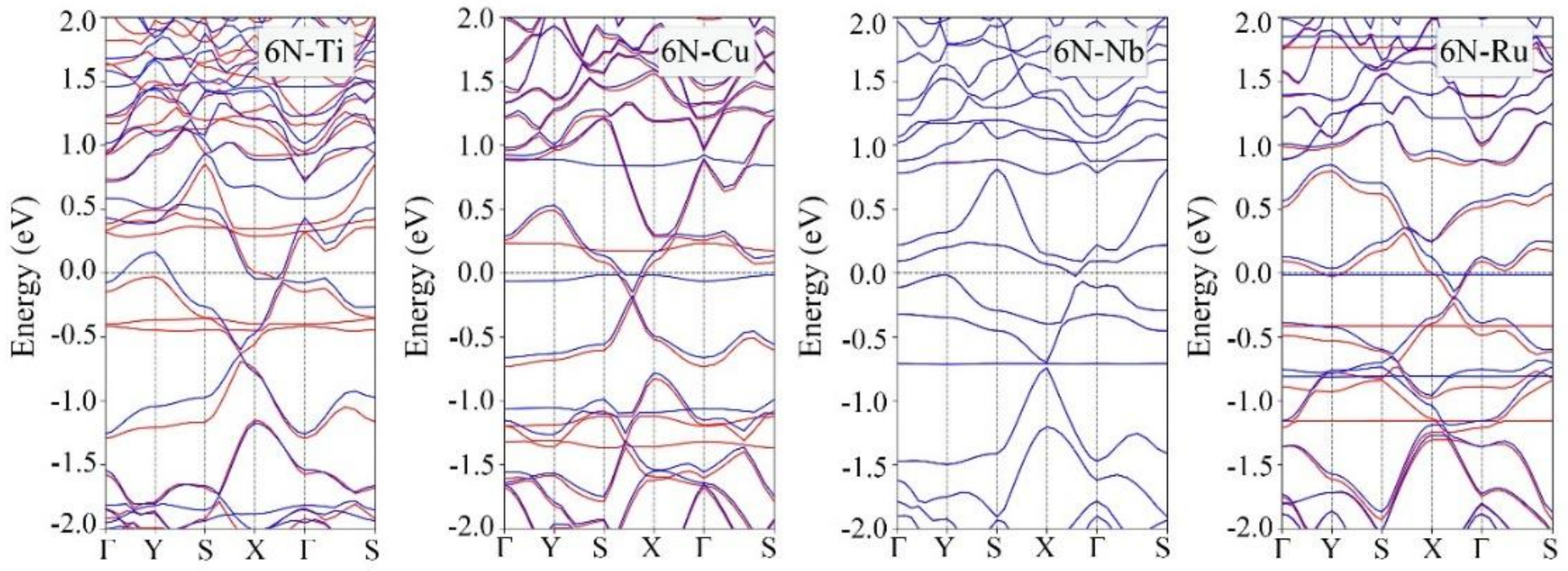

**Figure S13.** Band structures of $TM_2$@6N-Gr (TM = Ti, Cu, Nb and Ru). The red and blue lines represent spin-up and spin-down bands, respectively, and the Fermi level is set to 0 eV.